\documentclass[aps,prd,twocolumn,showpacs,superscriptaddress,nofootinbib,preprintnumbers]{revtex4-2}

\usepackage{style}

\begin{document}

\preprint{FERMILAB-PUB-23-097-T}

\title{Absorption of Vector Dark Matter Beyond Kinetic Mixing}

\author{Gordan Krnjaic\,\orcidlink{0000-0001-7420-9577}} 
\email{krnjaicg@fnal.gov}
\affiliation{Theoretical Physics Department, Fermi National Accelerator Laboratory, Batavia, Illinois 60510}
\affiliation{Department of Astronomy and Astrophysics, University of Chicago, Chicago, IL 60637}
\affiliation{Kavli Institute for Cosmological Physics, University of Chicago, Chicago, IL 60637}

\author{Tanner Trickle\,\orcidlink{0000-0003-1371-4988}}
\email{ttrickle@fnal.gov}
\affiliation{Theoretical Physics Department, Fermi National Accelerator Laboratory, Batavia, Illinois 60510}

\date{\today}

\begin{abstract}
Massive vector particles are minimal dark matter candidates that motivate a wide range of laboratory searches, primarily exploiting a postulated kinetic mixing with the photon. However, depending on the high energy field content, the dominant vector dark matter (VDM) coupling to visible particles may arise at higher operator dimension, motivating efforts to predict direct detection rates for more general interactions. Here we present the first calculation of VDM absorption through its coupling to electron electric (EDM) or magnetic (MDM) dipole moments, which can be realized in minimal extensions to the Standard Model and yield the observed abundance through a variety of mechanisms across the eV\,-\,MeV mass range. We compute the absorption rate of the MDM and EDM models for a general target, and then derive direct detection constraints from targets currently in use: Si and Ge crystals and Xe and Ar atoms. We find that current experiments are already sensitive to VDM parameter space corresponding to a cosmological freeze-in scenario, and future experiments will be able to completely exclude MDM and EDM freeze-in models with reheat temperatures below the electroweak scale. Additionally, we find that while constraints on the MDM interaction can be related to constraints on axion-like particles, the same is not true for the EDM model, so the latter absorption rate must be computed from first principles. To achieve this, we update the publicly available program~\textsf{EXCEED-DM} to perform these new calculations.
\end{abstract}

\maketitle

\section{Introduction}
\label{sec:intro}

Light vector particles are economical extensions to the Standard Model (SM) that require no stabilizing symmetries or mediator particles to account for the dark matter (DM) in our universe. The cosmological abundance of vector DM (VDM) can arise through a variety of mechanisms~\cite{Agrawal:2018vin,Antypas:2022asj,Arias:2012az,BasteroGil:2018uel,Co:2021rhi,Co:2018lka,Dror:2018pdh,Long:2019lwl,Nelson:2011sf,Adshead:2023qiw}. Minimally, VDM can be produced gravitationally through quantum fluctuations during inflation~\cite{Graham:2015rva}. Alternatively, the VDM abundance can 
arise through its SM interactions via the ``freeze-in" mechanism~\cite{Dodelson:1993je,Hall:2009bx} which relates VDM production at early times to observable signatures in terrestrial laboratories.

The most commonly studied VDM interaction is kinetic mixing with the SM photon through the $V_{\mu\nu} F^{\mu\nu}$ operator, where $V_\mu$ is the VDM field, with mass $m_V$, and $V_{\mu \nu} \equiv \partial_\mu V_\nu  - \partial_\nu V_\mu$ is its field strength tensor. This interaction can populate the VDM through the infrared (IR) freeze-in mechanism, in which the DM is initially absent at reheating and builds up through sub-Hubble interactions as the universe expands. While this mechanism is elegant and predictive, it is excluded for nearly all $m_V$ by a combination of direct and indirect detection searches~\cite{Pospelov:2008jk}, so there is motivation to explore alternative possibilities.

In the absence of kinetic mixing, the leading, viable, $V$\,-\,SM interactions are the electric (EDM) and magnetic dipole moment (MDM) operators,
\begin{align}
    \mathcal{L}_\text{MDM} &= \frac{d_M}{2} V_{\mu \nu} \bar{\Psi} \sigma^{\mu \nu} \Psi \label{eq:L_mdm} \\
    \mathcal{L}_\text{EDM} &= \frac{d_E}{2} V_{\mu \nu} \bar{\Psi} \, i\sigma^{\mu \nu} \gamma^5 \Psi \, , \label{eq:L_edm}
\end{align}
which have mass dimension five, $\sigma^{\mu\nu} = \frac{i}{2}[\gamma^\mu,\gamma^\nu]$, and $\Psi$ is a charged SM fermion field represented as a Dirac spinor in the broken electroweak phase. Such operators can be the leading VDM interaction with SM particles if suitable new states are integrated out at energy scales above $E > 1/d_{E,M}$ (for a concrete model see Refs.~\cite{Dobrescu:2004wz,Barducci:2021egn}). 

The phenomenology of the MDM model was fist studied in Ref.~\cite{Krnjaic:2022wor}, where it was shown that VDM with $\text{keV} \lesssim m_V \lesssim \text{MeV}$ can viably freeze-in, while avoiding indirect detection, and warm DM constraints. Since the MDM operator has mass dimension five, the cosmological abundance depends on the reheat temperature, $T_{\rm RH}$. This UV sensitivity makes it possible to viably freeze-in VDM by leveraging the potentially large $T_\text{RH}$ to enhance cosmological production, while evading indirect detection constraints that exclude freeze-in through kinetic mixing.

\begin{figure*}
    \includegraphics[width=\textwidth]{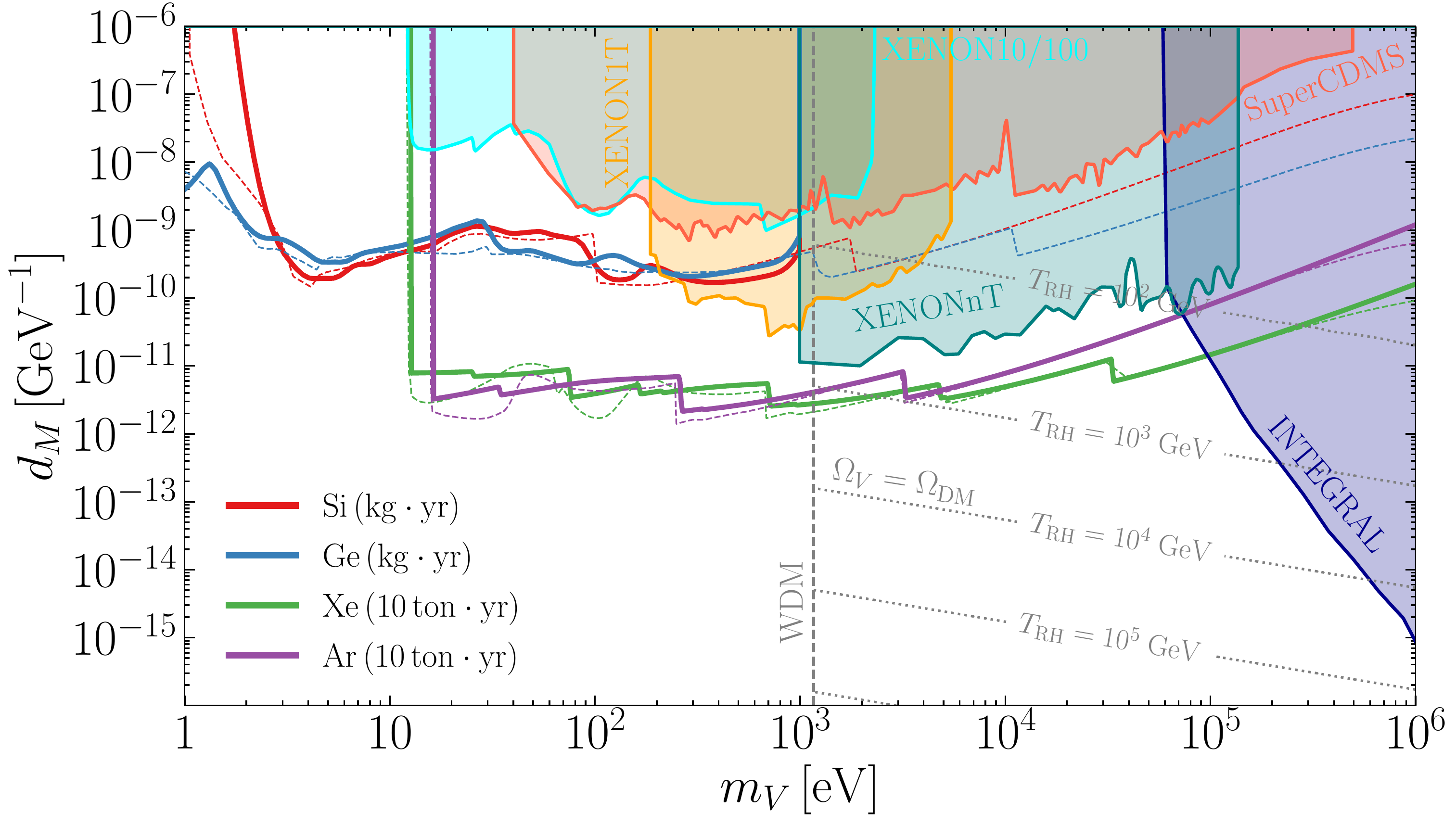}
    \caption{Projected 95\% C.L. constraints (3 events, no background) on the $d_M$ parameter in the MDM model, Eq.~\eqref{eq:L_mdm}, in crystal Si (red) and Ge (blue) targets, and atomic Xe (green) and Ar (purple) targets. Crystal targets assume an exposure of $1 \, \text{kg} \cdot \text{yr}$, and atomic targets assume an exposure of $10 \, \text{ton} \cdot \text{yr}$. Dashed lines are constraints using rescaled photon absorption data. Photon absorption data for the crystal targets is a combination of Refs.~\cite{1985, https://doi.org/10.18434/t48g6x}, and data for the atomic targets is a combination of Refs.~\cite{Henke_1993, https://doi.org/10.18434/t48g6x}, shown in Fig.~\ref{fig:photoelectric_cross_section_compare}. Previous constraints on ALPs from XENONnT~\cite{XENON:2022ltv}, XENON1T~\cite{XENON:2019gfn}, XENON10/100~\cite{Bloch:2016sjj}, and SuperCDMS~\cite{SuperCDMS:2019jxx} (shaded teal, orange, cyan, red, respectively) have been recast by converting those constraints to photon absorption cross sections, and using Eq.~\eqref{eq:rate_MDM}. Indirect detection bounds on $V \to 3 \gamma$ from INTEGRAL~\cite{Bouchet:2008rp,Krnjaic:2022wor} are shown in shaded blue. Gray dashed lines apply if the DM is produced via the freeze-in mechanism~\cite{Hall:2009bx, Krnjaic:2022wor}. The warm DM limit (WDM) is taken from Ref.~\cite{Krnjaic:2022wor}, and the lines labelled by $T_\text{RH}$ correspond to the necessary reheat temperature to generate the relic abundance via Eqs.~\eqref{eq:omega-pre-ewsb} and \eqref{eq:omega-post-ewsb}.}
    \label{fig:dM_reach}
\end{figure*}

While Ref.~\cite{Krnjaic:2022wor} mainly studied the indirect detection bounds on the MDM interaction, direct detection constraints were left for future work. In this paper, we extend this analysis to study VDM absorption onto atomic and crystal targets: 
\begin{itemize}
    \item{\bf Atomic Targets:} For $\text{keV} \lesssim m_V \lesssim \text{MeV}$, large exposure liquid noble experiments, e.g., XENON~\cite{XENON:2022ltv, XENON:2019gfn, Bloch:2016sjj} and DarkSide~\cite{DarkSide:2022knj, DarkSide:2018bpj}, are expected to be sensitive to absorption events when the VDM model couples to the electron. We will refer to these targets as ``atomic targets," since we approximate them as a collection of individual atoms, such that the total absorption rate is a simple sum of contributions from each atom. Atomic targets are especially interesting because they close the open window between the low mass ($m_V \sim \text{keV}$) warm DM constraints, and the indirect detection bounds which are dominant at higher ($m_V \sim$ MeV) masses. As we will see, these targets also play a key role in testing the predictive freeze-in scenarios for a wide range of $T_\text{RH}$.

    \item{\bf Crystal Targets:} For lower masses, $m_V \lesssim \text{keV}$, freeze-in is not a viable production mechanism since the DM would be too warm. However, there are a variety of production mechanisms which can populate the DM; see Sec.~\ref{subsec:production_from_additional_BSM} for more details. For these DM models, the energy levels of atomic targets are no longer suitable for efficient VDM absorption. Thus, for $\text{eV} \lesssim m_V \lesssim \text{keV}$ we also compute absorption rates and extract constraints for crystal Si and Ge targets with lower energy thresholds. These targets are utilized in several current and future experiments including, CDEX~\cite{CDEX:2022kcd}, DAMIC~\cite{DAMIC:2016qck,DAMIC:2019dcn,DAMIC:2020cut,DAMIC:2015znm,Settimo:2020cbq}, EDELWEISS~\cite{EDELWEISS:2019vjv,EDELWEISS:2018tde,EDELWEISS:2020fxc}, SENSEI~\cite{SENSEI:2019ibb,SENSEI:2020dpa,Crisler:2018gci}, and SuperCDMS~\cite{SuperCDMS:2018mne,CDMS:2009fba,SuperCDMS:2020ymb}. 
\end{itemize}

\begin{figure*}
    \centering
    \includegraphics[width=\textwidth]{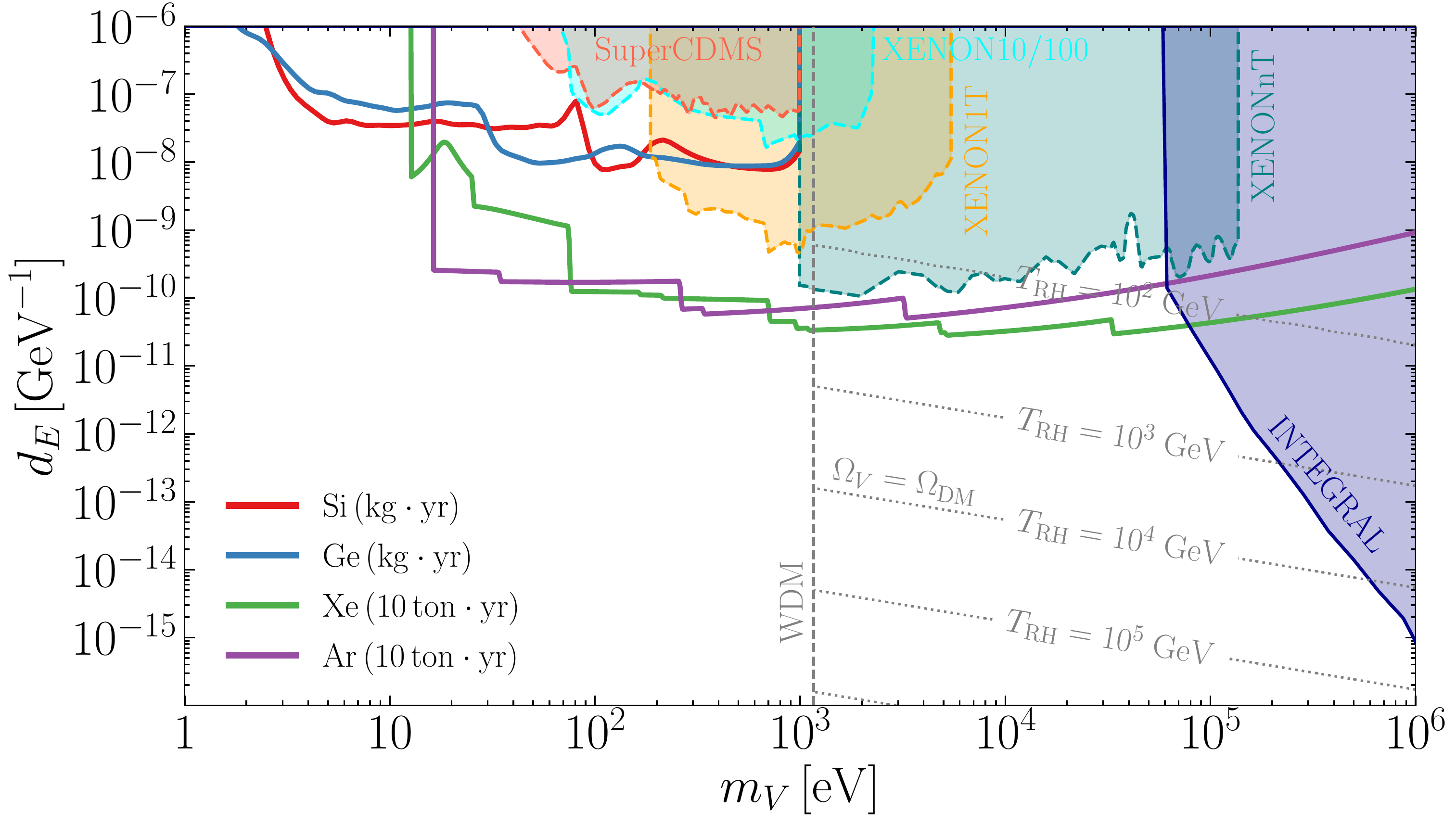}
    \caption{Projected 95\% C.L. constraints (3 events, no background) on the $d_E$ parameter in the EDM model, Eq.~\eqref{eq:L_edm}, in crystal Si (red) and Ge (blue) targets, and atomic Xe (green) and Ar (purple) targets. Crystal targets assume an exposure of $1 \, \text{kg} \cdot \text{yr}$, and atomic targets assume an exposure of $10 \, \text{ton} \cdot \text{yr}$. Shaded regions with dashed outline are expected direct detection constraints derived by rescaling the constraints in analogy with Fig.~\ref{fig:dM_reach}. Indirect detection bounds from INTEGRAL~\cite{Bouchet:2008rp} are shown in shaded blue. Gray dashed lines apply if the DM is produced via the freeze-in mechanism~\cite{Hall:2009bx, Krnjaic:2022wor}. The warm DM limit is taken from Ref.~\cite{Krnjaic:2022wor}, and the lines labelled by $T_\text{RH}$ correspond to the reheat temperature that yields the observed VDM abundance via Eqs.~\eqref{eq:omega-pre-ewsb} and \eqref{eq:omega-post-ewsb}.}
    \label{fig:dE_reach}
\end{figure*}

 In principle $\Psi$, in Eqs.~\eqref{eq:L_mdm} and \eqref{eq:L_edm}, could be any charged SM fermion. However, since our main focus is VDM absorption onto direct detection targets, for the remainder of this work, we will only consider the electron coupling. Furthermore, throughout our analysis, we treat the EDM and MDM cases separately, though our results generalize easily to scenarios in which both $d_M$ and $d_E$ are nonzero. 

Previous VDM absorption rate calculations~\cite{Arisaka:2012pb,Derbin:2012yk,Pospelov:2008jk, Mitridate:2021ctr, Chen:2022pyd, Trickle:2022fwt} have been focused on the kinetically mixed scenario, for which the absorption rate is simply related to the photon absorption rate. However, this relation does not hold in general; for generic interactions, the absorption rate must be calculated from first principles~\cite{Mitridate:2021ctr, Chen:2022pyd}. For our operators of interest, Eqs.~\eqref{eq:L_mdm} and \eqref{eq:L_edm}, we generalize the procedure outlined in Ref.~\cite{Mitridate:2021ctr} and find that, while the MDM absorption rate can be related to the photon absorption rate, the same is not true for the EDM interaction, which requires a dedicated calculation.

Moreover, we show that for $100 \, \text{keV} \lesssim m_V$ even the absorption rate of the familiar kinetically mixed model cannot be related to the photon absorption rate due to the kinematic mismatch (see Sec.~\ref{sec:absorption_calculation}). We perform the first principles 
absorption calculation by extending the publicly available code \textsf{EXCEED-DM}~\GithubLink~\cite{exdmv1, Trickle:2022fwt, Griffin:2021znd} with support for atomic targets, and make the modifications publicly available as well.

This paper is organized as follows. In Sec.~\ref{sec:cosmo} we discuss cosmological production mechanisms that can populate VDM in the early universe. In Sec.~\ref{sec:absorption_calculation}
we derive the electronic absorption rate of VDM in the MDM and EDM models. In Sec.~\ref{sec:results} we begin by comparing our first principles calculation of the VDM absorption rate to previously computed photoelectric cross section, and then we compute, and discuss, the direct detection constraints for both the MDM and EDM models. Lastly, in Sec.~\ref{sec:conclusion} we summarize our results and discuss how future work may extend experimental sensitivity to these scenarios.

\section{Cosmological Production}
\label{sec:cosmo}

Viable DM candidates that allow for absorption processes are generically out of equilibrium in the early universe, as the interaction strengths required to thermalize with the SM also induce particle lifetimes much shorter than the age of the universe. Thus, in this section, we briefly survey a variety non-thermal VDM production mechanisms, categorized according to whether or not the abundance arises from SM interactions. 

\subsection{Production From Additional BSM Fields}
\label{subsec:production_from_additional_BSM}

In a broad class of models, the non-thermal VDM abundance
depends on the details of the very early 
universe. For example, if the $V$ mass is nonzero during 
inflation and light compared to the Hubble rate, there is a cosmic abundance
of longitudinally polarized vectors arising from inflationary fluctuations~\cite{Graham:2015rva}, 
\be
\Omega_V \simeq \Omega_{\rm DM} \brac{H_I}{10^{14} \, \rm GeV}^2 \brac{6 \,\mu \rm eV}{m_V}^{1/2} \, ,
\ee
where $H_I$ is the inflationary Hubble scale, $\Omega_i \equiv \rho_i/\rho_c$ is the present day abundance fraction of species $i$, 
$\rho_{\rm c} \approx 4 \times 10^{-47} \, \text{GeV}^4$ is the critical density,
and $\Omega_{\rm DM} = 0.264$~\cite{Planck:2018vyg}. 
If $V$ couples to additional fields that undergo non-trivial evolution in the early universe, the VDM production rate can be further be enhanced 
through parametric resonance, which can yield the observed DM abundance even if 
gravitational production through inflationary fluctuations is inefficient~\cite{Dror:2018pdh,Adshead:2023qiw}. 
The VDM abundance can also arise from initial conditions via 
pre-inflationary misalignmnent \cite{Nelson:2011sf,Arias:2012az,Dimopoulos:2006ms}. 
However, as noted in Ref.~\cite{Arias:2012az}, misalignment 
is inefficient at producing VDM abundance unless
 the vector is non-minimally coupled to gravity. 

Since these mechanisms populate VDM independently of its
coupling to  the SM, we remain agnostic to the UV details
of such scenarios. Throughout this work, we assume that one
of these mechanisms suffices to produce the abundance -- particularly
in the low mass ($m_V \lesssim$ keV) regime, where SM freeze-in (discussed below)
production is excluded by structure formation bounds on warm DM.

\subsection{Production From SM Freeze-In}

The freeze-in mechanism postulates that DM is initially not populated
when the SM radiation bath is created 
after inflation. Self-consistency requires the DM-SM interaction
 rate to be sub-Hubble at this time, so that DM does not thermalize with visible
 matter. In this class of models, the $V$ abundance is, 
\be
\label{eq:omegaV}
\Omega_{\rm V} \simeq \frac{m_{V} s_0}{\rho_{\rm c}} \int_{T_{\rm RH}}^{T_{\rm IR}} \frac{dT}{T}
\frac{\langle \Gamma_{V}  \rangle \, \bar{n}_{V} }{H s} \,  ,
\ee
where $\langle \Gamma_{V} \rangle$ is the thermally averaged VDM production rate, $\bar{n}_{V}$ is the number density the $V$ particles would have if they were in chemical equilibrium at temperature $T$, $H$ is the Hubble expansion rate during radiation domination, $s$ is the entropy density, and a zero subscript represents a present day quantity. The integration range in \Eq{eq:omegaV} spans from the reheat temperature $T_{\rm RH}$ to the temperature at which freeze-in production halts, which typically satisfies $T_{\rm IR} = {\rm max}(m_V,m_{\rm SM})$, where $m_{\rm SM}$ is the mass of the main SM species driving freeze-in production. 

\subsubsection{Excluding Renormalizable Freeze-In}

Massive vectors that kinetically mix with the photon can be frozen in through this same interaction,
while maintaining a cosmologically long lifetime, for couplings that yield 
the observed DM abundance. For $m_V < 2m_e$, the dominant decay channel is 
$V\to 3\gamma$, which can be cosmologically metastable due to the sharp phase space suppression in the width for this process $(\propto m_V^9/m_e^8)$. However, the $\text{keV} \lesssim m_V \lesssim 2 m_e$ window is almost fully excluded by a combination
of X-ray and direct detection limits~\cite{Pospelov:2008jk}. For lighter ($m_V \lesssim$ keV) masses, these direct bounds
can be evaded, but in this regime the VDM is too warm for viable structure formation. 

If the vector particle is the gauge boson of an
abelian SM extension -- e.g., gauged $B-L$ or $L_i - L_j$, where $B$ and $L$ are the
baryon and lepton number, respectively -- 
it can couple directly to visible particles in the absence
of kinetic mixing (see Ref.~\cite{Bauer:2018onh} for a review). However,
in the absence of additional new field content
at low energies, all anomaly free $U(1)$ extensions require $V$ couplings
to neutrinos. Thus, for gauge couplings that would produce the observed DM 
abundance (e.g., $g \sim 10^{-11}$ for the ALP and kinetically mixed dark photon models~\cite{Pospelov:2008jk}), $V \to \bar \nu \nu$ decays are prompt
on cosmological timescales, so the vector is not a viable DM candidate. 

Similar considerations apply to VDM whose population freezes in through other dimension four operators (e.g., $V_\mu \bar \Psi \gamma^\mu \gamma^5 \Psi$). Since IR dominated freeze-in predicts a one-to-one correspondence between production and late time decay, the irreducible loop-level decay $V\to 3\gamma$ decay is comparably constrained by the same X-ray bounds that exclude kinetic mixing. Thus, any viable model of VDM freeze-in through SM interactions must involve operators beyond mass dimension four. 

\subsubsection{Freeze-In Through Dipole Operators}

In light of the above considerations, we now consider freezing in
VDM through MDM and EDM interactions. In the early universe, if electroweak symmetry is restored at high temperatures, the interactions in Eqs.~\eqref{eq:L_mdm} and \eqref{eq:L_edm} can be resolved as, 
\be
\label{eq:pre-ewsb-mdm}
{\cal L}^\text{UV}_\text{MDM} &=& \frac{d_M}{\sqrt{2}v} {\cal H}  
V_{\mu\nu}
\bar L  \sigma^{\mu \nu} 
e^c
\\
\label{eq:pre-ewsb-edm}
{\cal L}^\text{UV}_\text{EDM} &=& \frac{i d_E}{\sqrt{2}v} {\cal H}  
V_{\mu\nu}
\bar L  \sigma^{\mu \nu} \gamma^5 
e^c,
\ee
where $L$ is the first generation lepton doublet, $e^c$ is the 
right handed electron singlet, and $\cal H$ is the Higgs doublet. After electroweak symmetry breaking (EWSB), the Higgs doublet acquires a vacuum expectation value
$\langle {\cal H} \rangle = [ 0,  v/\sqrt{2}]^T$, where $v = 246$ GeV, and the 
operators in Eqs. \eqref{eq:pre-ewsb-mdm} and \eqref{eq:pre-ewsb-edm} reduce
to the MDM and EDM dipole interactions in Eqs.~\eqref{eq:L_mdm} and \eqref{eq:L_edm}, respectively. 

Since these interactions are higher dimension operators, the VDM freeze-in 
abundance will depend on the reheat temperature of the universe after inflation, 
$T_{\rm RH}$. If $T_{\rm RH} > 160$ GeV then the universe is initially in the unbroken
electroweak phase~\cite{DOnofrio:2015gop}, and the freeze-in abundance accumulates through the interactions in Eqs.~\eqref{eq:pre-ewsb-mdm} and \eqref{eq:pre-ewsb-edm}. As shown in Ref.~\cite{Krnjaic:2022wor}, this yields,
\be
\label{eq:omega-pre-ewsb}
\Omega_V \approx \Omega_{\rm DM} \brac{m_V}{3 \rm \, MeV}
\brac{d_{M/E} \cdot \rm GeV}{10^{-13}}^2 \brac{T_{\rm RH}}{\rm TeV}^3 \, ,
\ee
where VDM is produced via $e h \to  e V$ and $e^+ e^- \to h V$ reactions, and $h$ is the Higgs field. Note that Eq.~\eqref{eq:omega-pre-ewsb} holds for both the MDM and EDM models.

If, on the other hand, $T_{\rm RH} < 160$ GeV, the radiation era begins in the broken electroweak phase, and VDM freeze-in proceeds through the interactions in Eqs. \eqref{eq:L_mdm} and \eqref{eq:L_edm}, for which the abundance satisfies \cite{Krnjaic:2022wor},
\be
\label{eq:omega-post-ewsb}
\Omega_V \approx \Omega_{\rm DM} \brac{m_V}{ \rm MeV}
\brac{d_{M/E} \cdot \rm GeV}{10^{-10}}^2 \brac{T_{\rm RH}}{\rm GeV} \, .
\ee
Here, the main freeze-in reactions are now $e \gamma \to e V$ and $e^+ e^- \to \gamma V$. The key difference between Eq.~\eqref{eq:omega-post-ewsb} and Eq.~\eqref{eq:omega-pre-ewsb} is the $T_\text{RH}$ dependence; the VDM abundance is more strongly dependent on $T_\text{RH}$ in the unbroken electroweak phase. Note that in both Eqs. \eqref{eq:omega-pre-ewsb} and \eqref{eq:omega-post-ewsb}, we only use one of the MDM/EDM operators to calculate the abundance. Additionally, similar to Eq.~\eqref{eq:omega-pre-ewsb}, this expression holds for both the MDM and EDM models. In Figs.~\ref{fig:dM_reach} and \ref{fig:dE_reach} we show gray dashed contours corresponding to the $d_{M/E}$ necessary to generate the observed freeze-in abundance for various choices of $T_{\rm RH}$ and $m_V \gtrsim \text{keV}$. For smaller masses, freeze-in produces warm VDM in conflict with the observed matter power spectrum on small scales~\cite{Krnjaic:2022wor}, and therefore the abundance curves do not extend below $m_V \sim \text{keV}$. 

As discussed earlier, for smaller masses the abundance must be set by other mechanisms. However, for this to be true, it must be the case that the VDM is not thermalized via pair annihilation and Compton-like scattering processes. Conservatively assuming $T_{\rm RH} = 1$ MeV leads to a cosmological consistency condition of $d_{M/E} \lesssim 10^{-6} \, \text{GeV}^{-1}$, so we do not plot above above this value in Figs.~\ref{fig:dM_reach} and \ref{fig:dE_reach}.

\section{Absorption Rate Calculation}
\label{sec:absorption_calculation}

In this section we calculate VDM absorption rates through the MDM and EDM interactions in Eqs.~\eqref{eq:L_mdm} and \eqref{eq:L_edm}, respectively. Bosonic DM absorption rates in atomic targets have previously been calculated for the axion-like particles (ALP) and kinetically mixed dark photons~\cite{Arisaka:2012pb,Derbin:2012yk,Pospelov:2008jk}. However, since we are studying different DM interactions, the corresponding absorption rates need to be derived from first principles. 

In our calculation, we follow the approach in Refs.~\cite{Chen:2022pyd,Mitridate:2021ctr}, which extracted general absorption rates in terms of bosonic self-energies in the non-relativistic (NR) limit of the interaction Lagrangian.\footnote{The NR limit is appropriate here since the energy and momentum transfers in the process are both much smaller than the electron mass.} The advantages of this approach are that it applies to any DM model or target electronic structure, and automatically incorporates any in-medium effects (although these are mainly important for crystal targets with $\mathcal{O}(\text{eV})$ band gaps).

If the dark photon, $V$, does not mix with the photon, $A$, the optical theorem tells us that the absorption rate of the $i^\text{th}$ polarization of $V$, $\Gamma_V^i$, is given by,
\begin{align}
    \Gamma^i_V = - \frac{1}{m_V} \text{Im} \left[ \Pi_{VV}^i \right] \, ,
\end{align}
where $\Pi_{VV}^i$ is the self-energy of the $i^\text{th}$ polarization.
However, if $V$ mixes with $A$ (e.g., through a loop of electrons) then $V$ and $A$ are no longer the eigenstates of the theory, and the true eigenstates, $V'$ and $A'$, are those that diagonalize the $2\times2$ self-energy matrix between $V$ and $A$~\cite{Mitridate:2021ctr, Chen:2022pyd, Hardy:2016kme}. In this case, the VDM absorption rate is related to the imaginary part of the $V', V'$ self-energy,
\begin{align}
    \Gamma_{V'}^i & = -\frac{1}{m_V} \text{Im} \left[ \Pi_{V' V'}^i \right] \\
    & \approx - \frac{1}{m_V} \text{Im} \left[ \Pi_{VV}^i + \sum_j \frac{\Pi_{VA}^{ij} \Pi_{AV}^{ji}}{m_V^2 - \Pi_{AA}^j} \right] \, ,
\end{align}
where we have assumed that $V, A$ are perturbatively coupled. The absorption rate per unit exposure, averaged over the incoming DM polarizations, now becomes,
\begin{align}
    R = \frac{\rho_V}{\rho_T m_V} \frac{1}{3} \sum_{i = 1}^3 \Gamma_{V'}^i \, , \label{eq:intermediate_rate}
\end{align}
where $\rho_V = 0.4 \, \text{GeV} / \, \text{cm}^{3}$ is the DM mass density, and $\rho_T$ is the target mass density. Assuming that the self-energies are independent of polarization, shown explicitly in App.~\ref{app:self_energy_details} for the isotropic targets of interest here, Eq.~\eqref{eq:intermediate_rate} becomes,
\begin{align}
    R = - \frac{\rho_V}{\rho_T m_V^2} \text{Im} \left[ \Pi_{VV} + \frac{\Pi_{VA} \Pi_{AV}}{m_V^2 - \Pi_{AA}}\right] \, , \label{eq:abs_rate_self_energies}
\end{align}
and computing the absorption rate becomes a problem of evaluating the relevant self-energies, $\Pi_{VV}, \Pi_{VA}, \Pi_{AV}, \Pi_{AA}$. 

To calculate the self-energies we use an NR effective field theory (EFT) of electrons appropriate for energy and momentum transfers below the electron mass. This involves taking the NR limit of the QED Lagrangian, supplemented with the interaction terms from Eqs.~\eqref{eq:L_mdm} and \eqref{eq:L_edm}. This procedure is a tedious, but straightforward, exercise performed in Ref.~\cite{Mitridate:2021ctr}, and which we detail in App.~\ref{app:nr_op_via_FW}. The full expressions of the NR limit of the MDM and EDM Lagrangians, to $\mathcal{O}(m_e^{-2})$, can be found in Eqs.~\eqref{eq:L_NR_MDM_app} and \eqref{eq:L_NR_EDM_app}, respectively. 

While the full NR Lagrangians are relatively complicated, different approximations only leave a few important terms. First, we assume that the target has no spin ordering (i.e. there is no net electronic spin polarization) and that the electronic states are spin degenerate. This allows all of the self-energies to be written in terms of spin-independent matrix elements. Second, in typical targets the electron velocity, $v_e \sim Z \alpha \gtrsim 10^{-2}$, is greater than the halo DM velocity of $\sim 10^{-3}$. This allows us to neglect many terms which are proportional to the DM momentum, $\mathbf{q}$.

Explicitly, the terms which will give the dominant contribution to the absorption rates, via $\Pi_{VV}$, are,
\begin{align}
    \mathcal{L}^\text{NR}_\text{MDM} & \supset \frac{i d_M m_V }{m_e} \psi^\dagger \left( \bm{\sigma} \times \mathbf{k} \right) \psi \cdot \mathbf{V} \label{eq:L_NR_MDM_LO} \\ 
    \mathcal{L}^\text{NR}_\text{EDM} & \supset \frac{i d_E m_V }{m_e^2} \psi^\dagger \left( \bm{\sigma} \cdot \mathbf{k} \right) \mathbf{k} \, \psi \cdot \mathbf{V} \label{eq:L_NR_EDM_LO} \, ,
\end{align}
for the MDM and EDM models, respectively, where $\bm{\sigma}$ are the Pauli matrices, and $\mathbf{k} = -i \nabla$ is the electron momentum. These lead to the self-energies,
\begin{align}
    \Pi_{VV}^\text{MDM} & = \frac{2}{3} d_M^2 \omega^2 \, \bPi_{v^i, v^i} \label{eq:MDM_VV} \\
    \Pi_{VV}^\text{EDM} & = \frac{d_E^2 \omega^2}{3} \, \bPi_{v^i v^j, v^i v^j} \label{eq:EDM_VV}
\end{align}
where $i, j$ are summed indices, $\omega$ is the energy flowing through the self-energy diagram, and $v^i$ are the components of $\mathbf{v} \equiv \mathbf{k}/m_e$. The $\bar{\Pi}_{\mathcal{O}_1, \mathcal{O}_2}$ are then computed in terms of the target electronic structure~\cite{Mitridate:2021ctr, Chen:2022pyd, Trickle:2022fwt},
\begin{align}
    \bar{\Pi}_{\mathcal{O}_1, \mathcal{O}_2} = \frac{1}{V} \sum_{IF} \frac{1}{\langle F | F \rangle} \left[ \frac{\mathcal{T}_{\mathcal{O}_1}\mathcal{T}_{\mathcal{O}_2}^*}{\omega - \Delta \omega + i \delta} - \frac{\mathcal{T}_{\mathcal{O}_2}\mathcal{T}_{\mathcal{O}_1}^*}{\omega + \Delta \omega - i \delta} \right] \label{eq:reduced_self_energy}
\end{align}
where $V$ is the target volume, $| I \rangle, | F \rangle$ are the initial and final electronic states, respectively, $\omega_I, \omega_F$ are the initial and final state energies, respectively, $\mathcal{T}_{\mathcal{O}} \equiv \langle F | \mathcal{O} | I \rangle$ is the transition 
matrix element for Hermitian operator $\cal O$, and $\delta$ is the width of the electron resonance.\footnote{Strictly speaking, there should be an additional phase factor in the definition of the transition matrix element: $\mathcal{T}_\mathcal{O} = \langle F | e^{i \mathbf{q} \cdot \mathbf{x}} \mathcal{O}| I \rangle$~\cite{Mitridate:2021ctr, Chen:2022pyd, Trickle:2022fwt}. However, for absorption kinematics, $q \ll \omega$, the phase factor is generally negligible, except for $\mathcal{O} = 1$ due to state orthonormality. Therefore when discussing $\mathcal{T}_1$ we keep the phase factor.} This expression will take different forms in crystal and atomic targets, since the electronic states, $| I \rangle, | F \rangle$, differ between them. Explicit forms
for the transition matrix elements that define $\bar \Pi_{{\cal O}_1, {\cal O }_2}$ for crystal targets have been discussed in detail in Refs.~\cite{Chen:2022pyd, Mitridate:2021ctr, Trickle:2022fwt}. In App.~\ref{app:self_energy_details} we derive the results for atomic targets. Note that this definition of $\bPi$ is a slight generalization from previous works to account for non-unit normalized final states, $\langle F | F \rangle \neq 1$. This is useful when working with a continuum of final states, as appropriate for atomic targets.

Additionally, one can show that starting from the complete Lagrangians in App.~\ref{app:nr_op_via_FW}, at leading order the $V, A$ mixing self-energies are only non-zero in the MDM model, and are related to the photon self-energy,
\begin{align}
    \Pi_{VA}^\text{MDM} & = \Pi_{AV}^\text{MDM} = - \frac{d_M}{e} \frac{\omega^2}{2m_e} \Pi_{AA} \, , \label{eq:MDM_VA}
\end{align}
so the MDM model generates $m_V/m_e$ suppressed mixing effects, while there are no mixing effects in the EDM model. 

With all of the self-energies computed, we can now compute the absorption rates for the MDM and EDM models by substituting Eqs.~\eqref{eq:MDM_VV},~\eqref{eq:EDM_VV} and \eqref{eq:MDM_VA} in to Eq.~\eqref{eq:abs_rate_self_energies}. While the expressions in terms of the self-energies are identical between different targets, for the MDM model the rate can be written in terms of the photon self-energy since both the $V, A$ mixing term and the imaginary part of the $V, V$ self-energy are related to $\Pi_{AA}$,
\begin{align}
    \frac{1}{3} \text{Im} \left[ \bPi_{v^i, v^i} \right] = \frac{1}{e^2} \text{Im}\left[ \Pi_{AA} \right] \, .
\end{align}

Therefore, while the MDM model absorption rate can be written in terms of the photon self-energy, that does \textit{not} necessarily imply that it is related to the photon absorption rate. The reason for this is kinematics: when a photon with energy $\omega$ is absorbed, the momentum absorbed by the target is $q = \omega$. Therefore the photon absorption rate is determined by $\Pi_{AA}(\mathbf{q} = \omega \hat{\mathbf{q}}, \omega)$, where $\hat{\mathbf{q}}$ is the direction of the incoming photon. However, when VDM with energy $\omega \approx m_V$ is absorbed, the momentum absorbed by the target is much smaller, $q \sim m_V v_V$, where $v_V \sim 10^{-3}$ is the VDM velocity. Therefore, only when,
\begin{align}
    \Pi_{AA}(\mathbf{q} = m_V \hat{\mathbf{q}}, m_V) \approx \Pi_{AA}( \mathbf{q} \rightarrow 0, m_V) \, , \label{eq:photon_absorb_approx}
\end{align}
can the VDM absorption rate be related to the photon absorption rate, at incoming photon energies of $\omega = m_V$. To understand when this is a good approximation, it is important to know that $\Pi_{AA}$ is a function of the matrix element $\mathcal{T}_1$. If the dipole approximation is 
valid,
\begin{align}
   {\cal T}_1 = \langle F | e^{i \mathbf{q} \cdot \mathbf{x}} | I \rangle \approx i \mathbf{q} \cdot \langle F | \mathbf{x} | I \rangle \, , \label{eq:dipole_approx}
\end{align}
then Eq.~\eqref{eq:photon_absorb_approx} is also approximately true, since the $\Pi_{AA}(\mathbf{q} \rightarrow 0, m_V)$ depends on the approximated, right side of Eq.~\eqref{eq:dipole_approx}, and $\Pi_{AA}(\mathbf{q} = m_V \hat{\mathbf{q}}, m_V)$ depends on the left side of Eq.~\eqref{eq:dipole_approx} evaluated at $\mathbf{q} = m_V \hat{\mathbf{q}}$. The dipole approximation is valid when $q x \ll 1$, where $x$ is a typical distance scale. For atomic targets, typical $x$ values are $a_0 / Z$, where $a_0$ is the Bohr radius, and $Z$ is the nuclear charge. Therefore for Eq.~\eqref{eq:photon_absorb_approx} to be valid, $m_V \lesssim Z/a_0 \sim 4 Z\, \text{keV}$. For the atomic targets of interest here, Xe and Ar, this implies that the absorption rate of VDM with mass $m_V$, can only be related to the photon absorption rate, at $\omega = m_V$, for $m_V \lesssim 100 \, \text{keV}$. Moreover, the most accurate VDM absorption rate calculation for $m_V \gtrsim 100 \, \text{keV}$ would be a first principles calculation done in the dipole approximation, as opposed to rescaling the photon absorption rate.

While this is an important conceptual point, it is also at the boundary of interesting VDM parameter space, since indirect detection constraints from $V \rightarrow 3 \gamma$ become important near $m_V \sim \text{MeV}$. Therefore, for the most interesting VDM masses, it is appropriate to relate the MDM model absorption rate to the photon absorption rate.

For crystal targets, the natural way to express this is to write $\Pi_{AA}$ is terms of the dielectric function, $\epsilon(\omega)$~\cite{Mitridate:2021ctr},
\begin{align}
    \Pi_{AA} = \omega^2 (1 - \varepsilon)
\end{align}
while for atomic targets it is more natural to use the photoelectric cross section, $\sigma_\text{pe}$. These are related by,
\begin{align}
    \omega \, \text{Im}\left[ \varepsilon \right] \equiv \sigma_1 = n_T \sigma_\text{pe} \, ,
\end{align}
where $n_T$ is the target number density. Using these relations the absorption rates are given by,
\begin{align}
    R_\text{MDM} & = \frac{2 \rho_V}{\rho_T} \frac{d_M^2}{e^2} m_V^2 \text{Im} \left[ \varepsilon \right] \left( 1 + \frac{m_V^2}{8m_e^2} \frac{1 - |\varepsilon|^2}{|\varepsilon|^2}\right) \\
    & \approx \frac{2 \rho_V}{m_T} \frac{d_M^2}{e^2} m_V \sigma_\text{pe} \label{eq:rate_MDM} \\
    R_\text{EDM} & = -\frac{d_E^2}{3} \frac{\rho_V}{\rho_T} \text{Im} \left[ \bPi_{v^i v^j, v^i v^j} \right] \, ,\label{eq:rate_EDM}
\end{align}
where $m_T$ is the mass of the target atom. Therefore we see that while the MDM absorption rate can be related to photon absorption in a target, this is \textit{not} true for the EDM model; a similar result was found for the scalar DM absorption model discussed in Ref.~\cite{Mitridate:2021ctr}. Therefore to make projections for the direct detection constraints on the EDM model, the absorption rate must be computed from first principles.

\section{Direct Detection Constraints}
\label{sec:results}

From the discussion in Sec.~\ref{sec:absorption_calculation}, we know that while the absorption rate of the MDM model can be related to photon absorption, via the dielectric in crystal targets or the photoelectric cross section in atomic targets, the EDM absorption rate must be computed from first principles. Since the first principles calculation relies on an assumption about the initial and final electronic states in the target, we begin by discussing the electronic configurations assumed for the crystal Si, Ge, and atomic Xe, and Ar targets used here. 

For Si and Ge targets we use the publicly available electronic configuration from Ref.~\cite{https://doi.org/10.5281/zenodo.7246141}, which has been used previously~\cite{Trickle:2022fwt} to compute the absorption rate for scalar, axion-like particle, and kinetically mixed dark photon models. Detailed information about the electronic configuration can be found here~\cite{exdm_webpage}, and a longer discussion regarding modelling of the electronic configuration in this way can be found in Refs.~\cite{Griffin:2021znd, Trickle:2022fwt}. The configurations use three different methods to approximate the electronic states. The deeply bound, ``core" (all orbitals inclusively below 2$p$ in Si and 3$d$ in Ge) states are assumed to be solutions to the Hamiltonian of an isolated atom, which are computed using the RHF method~\cite{HF}. These states are expanded in an STO basis, and tabulated values of the coefficients can be found in Ref.~\cite{Bunge_1993}. The states closer to the Fermi surface, including four valence bands below and 60 (82) conduction bands above in Si (Ge), are computed with density functional theory (DFT) methods, expanded in a Bloch basis with an $E_\text{cut} = 2 \, \text{keV}$, and are all-electron reconstructed. Lastly, the highest energy states, with energies between $60 \, \text{eV}$ and $1 \, \text{keV}$ above the Fermi surface are treated as free plane waves. Lastly, following the treatment in Refs.~\cite{Mitridate:2021ctr, Trickle:2022fwt}, we model the electron width as $\delta = 0.2 + 0.1 \, \omega$.

Relative to the electronic states in crystal targets, those in atomic targets are much simpler. This is because all electrons in the target are tightly bound to an individual atom, and therefore the electronic states can be determined in isolation of the other atoms in the target. Similar to the deeply bound, ``core" electron states in crystal targets, we use the results of an RHF calculation~\cite{Bunge_1993} for the initial electronic states. For the final states, we use the continuum solutions to the Hamiltonian with a $V = - Z/r$ potential, where $Z$ is the nuclear charge, sometimes known as ``Coloumb wave functions"~\cite{Catena:2019gfa,Peng_2010,Sabbatucci_2016,Tan:2021nif}, with $Z$ set by the binding energy of the initial electronic state.\footnote{The $Z_F$ parameter used in calculating the final states is assumed to be related to the binding energy; see App.~\ref{app:atomic_transition_matrix_elements} for more details.} This approximation for the initial and final electronic states has been used in previous studies of DM-electron interactions in Xe and Ar targets~\cite{Catena:2019gfa, Catena:2022fnk, DarkSide:2018ppu}. More details about the initial and final electronic states, and the conventions used here, can be found in App.~\ref{app:atomic_transition_matrix_elements}.

To compute the absorption rates in Si, Ge, Xe, and Ar targets we use \textsf{EXCEED-DM}~\GithubLink~\cite{exdmv1, Trickle:2022fwt, Griffin:2021znd}. While \textsf{EXCEED-DM} has been used extensively with the electronic configurations of Si and Ge, it did not previously support atomic targets. We added support for this class of targets, and have made these updates publicly available in a new version. Additionally, the MDM and EDM model absorption rate calculation in Si and Ge targets has also been added.

\begin{figure*}
    \includegraphics[width=\textwidth]{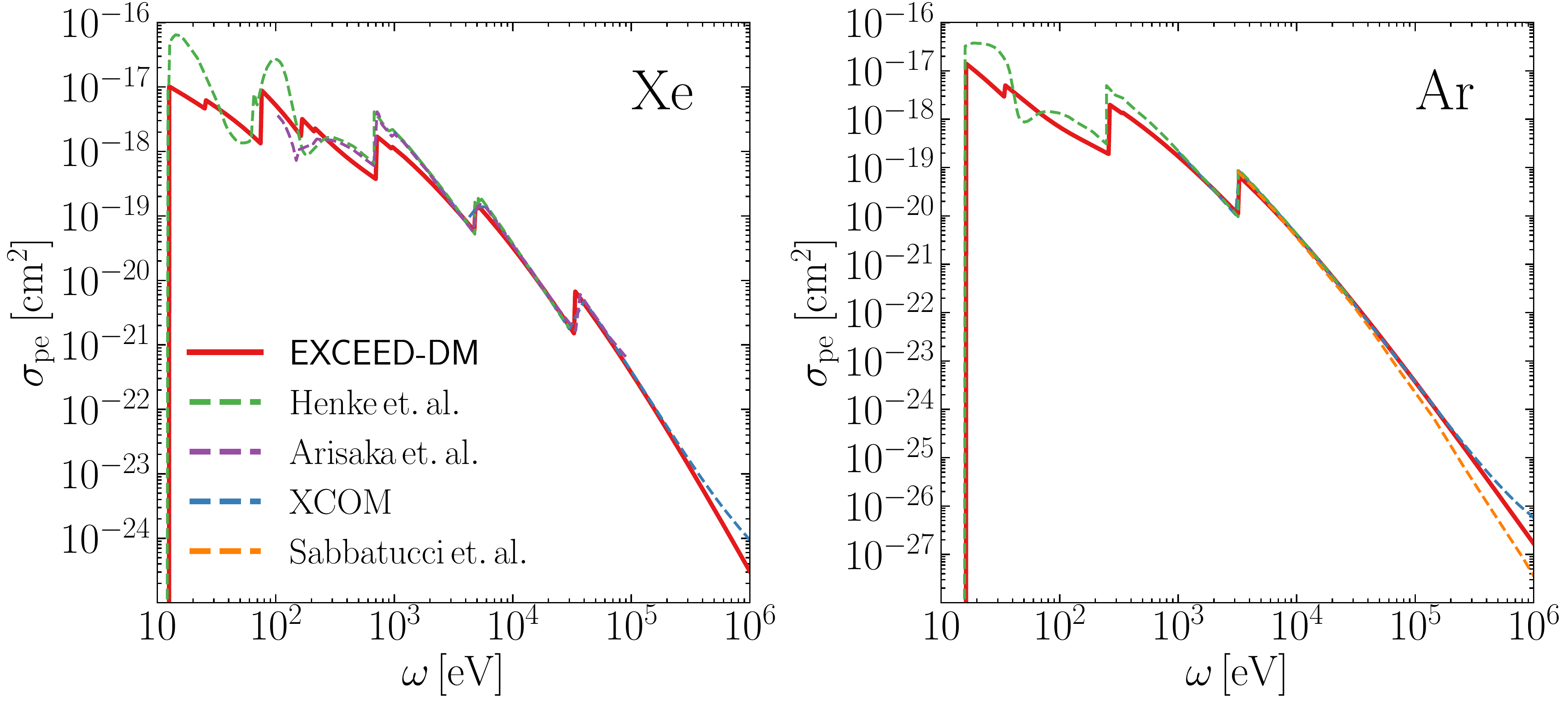}
    \caption{The photoelectric cross section, $\sigma_\text{pe}$, of Xe (left panel) and Ar (right      panel) targets computed with various methods. The calculation done with \textsf{EXCEED-DM}, and subsequently used to constrain the MDM model in Fig.~\ref{fig:dM_reach}, is shown in solid red. Experimental measurements from Ref.~\cite{Henke_1993} are shown in dashed green. Other experimental measurements, used in Ref.~\cite{Arisaka:2012pb}, to place constraints on other DM models with Xe targets, are shown in dashed purple. Photoelectric cross sections from the NIST database, computed with the XCOM program~\cite{https://doi.org/10.18434/t48g6x}, are shown in dashed blue. Lastly, we show the photoelectric cross section for the $K$ shell of Ar (dashed orange), computed under the dipole approximation in Ref.~\cite{Sabbatucci_2016}, to further illustrate the discrepancy between the photoelectric cross section computed with photon versus dark photon kinematics, as discussed in Sec.~\ref{sec:absorption_calculation}.}
    \label{fig:photoelectric_cross_section_compare}
\end{figure*}

Before discussing the constraints on the MDM and EDM models, it is important, when possible, to verify the electronic configurations being used against measured photon absorption data. This has been done previously for Si and Ge~\cite{Mitridate:2021ctr}, and therefore we focus on the Xe and Ar calculations. Using Eq.~\eqref{eq:rate_MDM} we can compute the photoelectric cross section, $\sigma_\text{pe}$, by rescaling the dark photon absorption rate in the MDM model. In Fig.~\ref{fig:photoelectric_cross_section_compare} we compare the photoelectric cross sections computed with \textsf{EXCEED-DM} to a variety of other calculations and measurements~\cite{Arisaka:2012pb,Henke_1993,https://doi.org/10.18434/t48g6x,Sabbatucci_2016}. Overall we find good agreement in both Xe and Ar targets, with $\mathcal{O}(1)$ discrepancies for $\omega \lesssim \, \text{keV}$, and $\omega \gtrsim 100 \, \text{keV}$. The slight disagreement at low energies is somewhat expected due to our ``isolated" atom approximation, which completely neglects the target environment when solving for the electronic wave functions. A more sophisticated approach (e.g., using the DFT formalism) for the electronic states involved in low $\omega$ absorption would likely reduce this discrepancy. 

At high energies, the difference between the XCOM and \textsf{EXCEED-DM} calculations is due to the kinematic difference between photon and dark photon absorption discussed in Sec.~\ref{sec:absorption_calculation}. This high energy discrepancy can also be understood in the context of the standard dipole approximation in Eq.~\eqref{eq:dipole_approx}. For photon absorption, when $q x = \omega x \gg 1$, or equivalently when $\omega \gtrsim Z/a_0,$ the dipole approximation breaks down, and one must use the exponential form of the operator in the transition matrix element. The XCOM calculation uses the exponential form, whereas the dashed orange curve from Ref.~\cite{Sabbatucci_2016} uses the dipole approximation, which underestimates the photon absorption rate at high $\omega$. Ref.~\cite{Sabbatucci_2016} also computes $\sigma_\text{pe}$ in the exponential form and reaches same conclusion: the dipole approximation underestimates the photon absorption rate for large $q = \omega$. However, while the dipole approximation is not appropriate for photon absorption at high $\omega$, it \textit{is} appropriate for dark photon absorption, since $q \ll \omega$, and therefore $q x \ll 1$ is still valid even when $\omega \sim \text{MeV}$, contrary to photon absorption.

With verification that our electronic configurations reproduce other observables, we now discuss the constraints for the MDM model, shown in Fig.~\ref{fig:dM_reach}, as well as the EDM model, shown in Fig.~\ref{fig:dE_reach}. For both the MDM and EDM models, the lowest probeable DM mass is set by the minimum energy difference between initial and final states, since $\omega \approx m_V$. For crystal targets this is the band gap, which is about $1.11 \, \text{eV}$ in Si, and $0.67 \, \text{eV}$ in Ge. In atomic targets this is the ionization energy (or negative of the binding energy) of the least bound electron, i.e., the 5p electron in Xe, with $E_I \approx -12 \, \text{eV}$ and the 3p electron in Ar, with $E_I \approx -16 \, \text{eV}$.\footnote{DM with a mass below the ionization energy could cause a transition from a filled bound state to an unfilled bound state in an atomic target. However, the observable would then be a low energy scintillated photon when the electron decays back down, as opposed to an outgoing electron, and therefore evade detection in the standard experimental detection channels.} The high DM mass cutoff in Si and Ge at $m_V \sim 1 \, \text{keV}$ is due to the fact that the electronic configuration only includes final states with final energies up to a keV. That is, the cutoff is just an analysis cutoff, not a physical one. However, for $m_V \gtrsim \text{keV}$ future iterations of the XENON experiments are expected to dominate the bounds due to their large exposure. Therefore it is unlikely that Si and Ge target projections will be important above for $m_V \gtrsim \text{keV}$.

As discussed in Sec.~\ref{sec:cosmo}, for $m_V \gtrsim \text{keV}$, the cosmological abundance can be set by the freeze-in mechanism. This lower bound is set by constraints on structure formation~\cite{Garzilli:2018jqh,Irsic:2017ixq,Zelko:2022tgf}, and is shown as a dashed gray vertical line in Figs.~\ref{fig:dM_reach},~\ref{fig:dE_reach}. If the DM was lighter than this it would be too hot, and suppress structure formation on small scales. For masses larger than this, the abundance is then set by the reheat temperature, $T_\text{RH}$, via Eqs.~\eqref{eq:omega-pre-ewsb} and \eqref{eq:omega-post-ewsb}. As shown in Fig.~\ref{fig:dM_reach}, we find that current direct detection constraints, mainly XENONnT~\cite{XENON:2022ltv}, set a lower bound on the viable reheat temperature for the MDM model, $T_\text{RH} \gtrsim 100 \, \text{GeV}$. The constraint increases to nearly $T_\text{RH} \gtrsim 10^3 \, \text{GeV}$ for $m_V \sim 1 \, \text{keV}$. This nearly closes the previously open parameter space on MDM VDM production via sub-electroweak scale reheat temperatures. Roughly $100 \times$ greater exposure will be needed to make the same statement about $T_\text{RH} \sim 10^3 \, \text{GeV}$, due to the stronger scaling of the abundance with $T_\text{RH}$ above the electroweak scale, as shown in Eq.~\eqref{eq:omega-pre-ewsb}. Furthermore, low threshold analyses of atomic targets, along with even lower threshold constraints from crystal target experiments such as CDEX~\cite{CDEX:2022kcd}, DAMIC~\cite{DAMIC:2016qck,DAMIC:2019dcn,DAMIC:2020cut,DAMIC:2015znm,Settimo:2020cbq}, EDELWEISS~\cite{EDELWEISS:2019vjv,EDELWEISS:2018tde,EDELWEISS:2020fxc}, SENSEI~\cite{SENSEI:2019ibb,SENSEI:2020dpa,Crisler:2018gci}, and SuperCDMS~\cite{SuperCDMS:2018mne,CDMS:2009fba,SuperCDMS:2020ymb} will constrain models producing low mass MDM coupled DM via the other mechanisms discussed in Sec.~\ref{sec:cosmo}.

Similar conclusions, to that of the MDM model, hold for the EDM model shown in Fig.~\ref{fig:dE_reach}; although the constraints are slightly weaker. This can be understood from Eqs.~\eqref{eq:rate_MDM},~\eqref{eq:rate_EDM}: the EDM absorption rate is suppressed relative to the MDM scenario by a factor of $v_e^2$, and therefore constraints are weaker by, roughly, a factor of $Z \alpha$. Since the EDM absorption rate is not related to the photon absorption rate, there are no official direct detection constraints. To get an estimate of where these would lie, we assume that the projections, solid lines in Fig.~\ref{fig:dE_reach}, can be rescaled by the same factor that would bring the solid lines in agreement with the edge of the shaded regions in Fig.~\ref{fig:dM_reach}. Essentially, we are assuming the same ``detection efficiency" in both the EDM and MDM models. These rescaled, expected limits are shown as shaded regions with dashed edges in Fig.~\ref{fig:dE_reach}.

\section{Conclusion}
\label{sec:conclusion}

In this paper we have studied the absorption rate of vector dark matter particles that couple to electrons preferentially through electric and magnetic dipole moment operators.
We use these results to place new limits on these scenarios from a variety of direct detection searches utilizing both crystal and atomic targets, and make
projections for future searches, which are poised to improve experimental sensitivity to these interactions by several orders of magnitude across 
the eV-MeV mass range. This parameter space is particularly interesting because it covers masses and dipole couplings that can 
yield predictive cosmological freeze-in production through the same operator responsible for absorption reactions; freeze-in
through kinetic mixing is nearly fully excluded for all choices of particle mass. 

While it has been known for some time that the absorption of the kinetically mixed dark photon, and ALPs, can be related to the photon absorption rate~\cite{Arisaka:2012pb,Derbin:2012yk,Pospelov:2008jk}, this relation does not hold in general. Indeed, in Ref.~\cite{Mitridate:2021ctr} it was shown that for scalar DM this relationship does not exist. Following the ideas presented in Ref.~\cite{Mitridate:2021ctr}, albeit with a different derivation discussed in App.~\ref{app:nr_op_via_FW}, in Sec.~\ref{sec:absorption_calculation} we derive the NR limit of UV Lagrangians in Eqs.~\eqref{eq:L_mdm},~\eqref{eq:L_edm} and use these to compute the self-energies and absorption rate. This procedure is general to any DM model, and therefore should be useful for future studies of different DM models. We find that while the absorption rate of the MDM model can be related to the photon absorption rate, this is not true for the EDM model, which therefore must be computed from first principles.

To compute the absorption rate of the EDM model, we modified \textsf{EXCEED-DM}~\GithubLink~\cite{exdmv1, Trickle:2022fwt, Griffin:2021znd}, the program previously used in the first principles study of DM absorption on crystal targets~\cite{Mitridate:2021ctr, Chen:2022pyd, Trickle:2022fwt}. We implemented two main improvements: first, we added support for absorption calculations involving atomic targets, i.e., transitions between bound and continuum states using the standard approximations for these electronic states~\cite{Catena:2019gfa, Catena:2022fnk}. Second, we added the EDM and MDM absorption rate calculation for crystal targets, e.g., Si and Ge (in addition to the atomic targets previously mentioned). Updates to the program are publicly available. 

The results of these new calculations were discussed in Sec.~\ref{sec:results}, and shown in Figs.~\ref{fig:dM_reach} and \ref{fig:dE_reach}. We began by verifying the first principles calculation against other photoelectric cross section measurements and calculations, in Xe and Ar targets, in Fig.~\ref{fig:photoelectric_cross_section_compare}. We find good agreement, up to $\mathcal{O}(1)$ factors, with the largest discrepancies below $\sim 100 \, \text{eV}$, and above $\sim 100 \, \text{keV}$. At low energies, the discrepancy is likely due to a too simplified treatment of the electronic states. Future work using more advanced methods, e.g., density functional theory, should decrease the disagreement and also lead to more accurate scattering rate calculations~\cite{Catena:2019gfa, Catena:2022fnk}. At high energies, the difference is due to using the dipole approximation in the transition matrix element. However, as discussed in Sec.~\ref{sec:absorption_calculation}, the dipole approximation is appropriate in the context of DM absorption, since kinematically the dark photon is depositing much less momentum, relative to a photon, for a given energy deposition. Therefore, strictly speaking, in this region even the kinetically mixed dark photon absorption rate cannot be related to the measured photon absorption rate. While this is a theoretically interesting point, the difference ends up being marginal; and, moreover, the DM masses for which this is important are ruled out by indirect detection, as seen in Figs.~\ref{fig:dM_reach} and \ref{fig:dE_reach}. 

VDM coupling to electrons via MDM and EDM operators are simple, motivated extensions to the SM which can account for the DM abundance, and therefore must be searched for in every possible avenue. In this paper, we have shown how to compute the direct detection rate for these models, and aim to include them in future official analyses alongside ALP and kinetically mixed dark photon constraints. Even without considering MDM and EDM couplings to other SM fermions, there is interesting physics beyond the scope of this paper yet to be explored. Searching for VDM masses below $m_V \sim \mathcal{O}(\text{eV})$ will require utilizing more novel excitations in low threshold experiments, such as phonons~\cite{Baym:2020uos,Caputo:2019cyg,Caputo:2020sys,Coskuner:2021qxo,Cox:2019cod,Griffin:2018bjn,Griffin:2020lgd,Griffin:2019mvc,Kahn:2020fef,Knapen:2017ekk,Kurinsky:2019pgb,Mitridate:2020kly,Schutz:2016tid,Trickle:2019nya,Trickle:2020oki} and magnons~\cite{Barbieri:1985cp,Chigusa:2020gfs,Esposito:2022bnu,Mitridate:2020kly,Trickle:2019ovy,Trickle:2020oki}, or electronic excitations in more novel targets, such as small band gap crystals~\cite{Chen:2022pyd,Coskuner:2019odd,Geilhufe:2019ndy,Hochberg:2017wce,Inzani:2020szg} and superconductors~\cite{Hochberg:2015fth,Hochberg:2016ajh,Hochberg:2015pha}. Additionally, a detailed study of stellar cooling constraints is important, and may place stronger constraints for $m_V \lesssim 10 \, \text{keV}$ than the thermalization requirement discussed in Sec.~\ref{sec:cosmo}.

\begin{acknowledgments}
We thank Bogdan Dobrescu, Simon Knapen, Duncan Rocha, and Anastasia Sokolenko for helpful conversations.
This work is supported by the Fermi Research Alliance, LLC under Contract No. DE-AC02-07CH11359 with the U.S. Department of Energy, Office of Science, Office of High Energy Physics.
The computations presented here were conducted on the Wilson-Institutional Cluster at Fermilab.
\end{acknowledgments}

\onecolumngrid
\appendix

\section{Non-Relativistic Lagrangians}
\label{app:nr_op_via_FW}

The initial step in any calculation of DM induced electronic excitation rates is to reduce the UV Lagrangian, written in terms of the four component Dirac field, $\Psi$, to the non-relativistic (NR) Lagrangian, written in terms of a light, two component field, $\psi$, which solves a charged particle Schr\"odinger equation. In general, this is a non-trivial problem for electrons in background electromagnetic potentials. To leading order in the DM-electron coupling, this can be done in two separate steps. The first step is to find the $\Psi \rightarrow \psi$ map which reduces to the NRQED Lagrangian starting from,
\begin{align}
    \mathcal{L}_\text{QED} = \bar{\Psi} \left( i \gamma^0 D_0 + i \gamma^i D_i - m_e \right) \Psi \, ,
    \label{eq:L_dirac}
\end{align}
where $D_\mu = \partial_\mu + i e A_\mu$. The second step is to apply the same field transformation, $\Psi \rightarrow \psi$, on a UV DM-electron interaction vertex, e.g., $\bar{\Psi} \mathcal{O}_\text{UV} \Psi$, which creates a UV to NR map to one written in terms of the light field, $\psi$,
\begin{align}
    \bar{\Psi} \mathcal{O}_\text{UV} \Psi \rightarrow \psi^\dagger \mathcal{O}_\text{NR} \psi \, ,
\end{align}
where $\mathcal{O}_\text{UV}$ is a $4 \times 4$ matrix, and $\mathcal{O}_\text{NR}$ is a $2 \times 2$ matrix.

In Sec.~\ref{subapp:nrqed_FW} we derive the NRQED Lagrangian to $\mathcal{O}(m_e^{-2})$. That is, we take the low energy, momentum limit of Eq.~\eqref{eq:L_dirac} to find $\mathcal{L}^\text{NR}_\text{QED}$, whose leading terms give the Schr\"odinger equation of a charged particle. While the NRQED Lagrangian has clearly been known for a long time, we rederive it starting from Eq.~\eqref{eq:L_dirac} since we are using a different method to take the UV to NR limit than was done in previous calculations~\cite{Mitridate:2021ctr}. Therefore a complete derivation is useful in comparing to previous results, as well as a pedagogical introduction to NRQED. Then, in Sec.~\ref{subapp:op_UV_to_NR}, we apply the field transformation on the MDM and EDM interactions of interest in Eqs.~\eqref{eq:L_mdm} and \eqref{eq:L_edm}, respectively. The final result of these calculations is summarized in Table~\ref{tab:nr_limit}.

\subsection{NRQED Lagrangian via Foldy-Wouthuysen Transformation}
\label{subapp:nrqed_FW}

We begin with the NRQED step, which can be done in many ways~\cite{Paz:2015uga}. One detailed recipe for doing this in the context of DM absorption on electrons is given in Ref.~\cite{Mitridate:2021ctr}. Here we use an alternative formulation, known in other contexts as a Foldy-Wouthuysen (FW) transformation~\cite{Balk:1993ev,Foldy_1952,Foldy:1949wa,Gardestig:2007mk,bj}. For this problem, and for reasons that will become clear shortly, it is better to work in the Dirac basis, where,
\begin{align}
    \gamma^0 = \begin{pmatrix} 1 & 0 \\ 0 & -1\end{pmatrix} \, , \, \gamma^i = \begin{pmatrix} 0 & \sigma^i \\ - \sigma^i & 0 \end{pmatrix} \, , \gamma^5 & = \begin{pmatrix} 0 & 1 \\ 1 & 0 \end{pmatrix} \, .
\end{align}
The main problem that we are trying to solve is that the QED Lagrangian in Eq.~\eqref{eq:L_dirac} mixes the two, two-component fields inside $\Psi$, which makes solving the system for each two-component field difficult. To see this define, 
\begin{align}
    \Psi = e^{-i m_e t} \begin{pmatrix} \psi \\ \psi_h \end{pmatrix} \, ,
\end{align}
where $\psi, \psi_h$ are two component fields, and substitute $\Psi$ back in to the QED Lagrangian,
\begin{align}
    \mathcal{L}_\text{QED}(\psi, \psi_h) = \psi^\dagger \left( i D_0 \right) \psi + \psi_h^\dagger \left( i D_0 - 2 m_e \right) \psi_h + \psi^\dagger_h \sigma^i D_i \psi + \psi^\dagger \sigma^i D_i \psi_h \, . \label{eq:L_QED_heavy_light}
\end{align}
Only $\psi_h$ has a mass term, and therefore it is referred to as the ``heavy" component, and $\psi$ is the ``light" component. We also see that $\psi$ and $\psi_h$ are coupled due to the presence of $\sigma^i D_i$. One approach to decouple the terms is to integrate out the heavy field, $\psi_h$. This is the approach taken in Ref.~\cite{Mitridate:2021ctr}. The idea here, and that of the FW procedure, is to perform consecutive field redefinitions, at each order in $1/m_e$,
\begin{align}
    \Psi \rightarrow e^{-i m_e t} U_1 U_2 \ldots \Psi \, ,\label{eq:FW_field_redef_schematic}
\end{align}
where $U_i$ are some operators acting on $\Psi$, to remove the mixing between $\psi$ and $\psi_h$, defined (in the Dirac basis) as the upper and lower components, respectively, of $\Psi$ on the right hand side of Eq.~\eqref{eq:FW_field_redef_schematic}. To do this efficiently it is important to identify the terms which mix $\psi$ and $\psi_h$, or equivalently, the upper and lower components of $\Psi$. 

In the context of FW transformations, the operators which mix $\psi$ and $\psi_h$ are known as \textit{odd} operators, and diagonal operators are \textit{even}. They are defined by their (anti) commutation relations with $\gamma^0$. Specifically, an odd operator, $\mathcal{O}$, satisfies, $\{ \mathcal{O}, \gamma^0 \} = 0$, and an even operator, $\mathcal{E}$, satisfies $[\mathcal{E}, \gamma^0] = 0$. The goal of the FW transformation is then to remove all odd operators from the Lagrangian at each order in $1/m_e$, thereby decoupling $\psi$ and $\psi_h$ at any given order. 

 Specifically, the recipe is to find $n$ Hermitian operators, $\{ X_0, \ldots, X_{n - 1} \}$, such that the field redefinition,
\begin{align}
    \Psi \rightarrow e^{-i m_e t} \left[ \exp\left( -i \frac{X_0}{m_e} \right)  \ldots \exp\left( -i \frac{X_{n - 1}}{m_e^n} \right) \right] \Psi \, ,\label{eq:FW_field_redef}
\end{align}
removes all the odd operators to $\mathcal{O}(m_e^{-n})$. To expand the QED Lagrangian, Eq.~\eqref{eq:L_dirac}, to ${\cal O}(m_e^{-2})$ we need to find $X_0, X_1$. One can show that the $X_0, X_1$ which do this are,
\be
    X_0 & = - \frac{1}{2} \gamma^i D_i ~~~,~~~
    X_1 & = \frac{e}{4} \gamma^0 \gamma^i F_{0i} \, , \label{eq:X0_X1}
\ee
where $[D_\mu, D_\nu] = ie F_{\mu \nu}$. With $X_0, X_1$ in hand the NRQED Lagrangian of $\psi$ to $\mathcal{O}(m_e^{-2})$ can be derived by substituting the $X_0, X_1$ in Eq.~\eqref{eq:X0_X1} to $\Psi$ in Eq.~\eqref{eq:FW_field_redef}, the $\Psi$ in to the QED Lagrangian in Eq.~\eqref{eq:L_dirac}. While seemingly inconspicuous, the phase factor $e^{-i m_e t}$ plays an important role here. To see this explicitly, note that under the FW transformation,
\be
  {\cal L}_\text{QED} &\to&  \Psi^\dagger  \left[ \exp\left( i \frac{X_1}{m_e^2} \right) \exp\left( i \frac{X_0}{m_e} \right) \right] e^{i m_e t} \biggl( i \gamma^0 \slashed{D} - \gamma^0 m_e \biggr) e^{-i m_e t} \left[\exp\left( -i \frac{X_0}{m_e} \right) \exp\left( -i \frac{X_1}{m_e^2} \right) \right] \Psi \, .
\ee
Furthermore, we can define $P_- \equiv \frac{1}{2} (1 - \gamma^0)$ and rewrite the 
QED Lagrangian without the $e^{im_e t}$ phase as,
\be
   {\cal L}_\text{QED} = \Psi^\dagger  \left[ \exp\left( i \frac{X_1}{m_e^2} \right) \exp\left( i \frac{X_0}{m_e} \right) \right]  \biggl( i D_0 + i \gamma^0 \gamma^i D_i + 2 m_e P_- \biggr) \left[\exp\left( -i \frac{X_0}{m_e} \right) \exp\left( -i \frac{X_1}{m_e^2} \right) \right] \Psi  \label{eq:NRQED_FW_mid}
\ee
 and, again, $X_0, X_1$ are given by Eq.~\eqref{eq:X0_X1}. Therefore, as in the derivation of Eq.~\eqref{eq:L_QED_heavy_light}, this phase factor introduces an operator projecting the mass on to the lower component of $\Psi$. Following the derivation in Eq.~\eqref{eq:NRQED_FW_mid} further, the $\psi$ part of the QED Lagrangian becomes,
\begin{align}
    \mathcal{L}_\text{QED}^\text{NR} & = \psi^\dagger \, \text{Tr} \left[ P_+ \left( i D_0 - \frac{2}{m_e} \gamma^0 X_0^2 - \frac{1}{m_e^2} \gamma^0 \left\{ X_1, X_0 \right\} \right) \right] \psi + \mathcal{O} \left( \frac{1}{m_e^3}\right) \\
    & \supset 
        e \psi^\dagger \left[ 
            -1 + 
            \frac{1}{8 m_e^2} \left( \mathbf{p}^2 + 2i \mathbf{p} \cdot \cross{\bsig}{\mathbf{k}}{} \right) 
        \right] A_0 \psi 
        + e \psi^\dagger \left[ 
            \frac{1}{2m_e} \left( \mathbf{K}' + i \cross{\bsig}{\mathbf{p}}{} \right) 
            + \frac{\omega'}{8m_e^2} \left( -\mathbf{p} - i \cross{\bsig}{\mathbf{K}'}{} \right) 
        \right] \cdot \mathbf{A} \psi \nonumber \\
    & - \frac{e^2}{2m_e} \mathbf{A}^2 \psi^\dagger \psi 
    + \frac{ie^2}{4m_e^2} \psi^\dagger \left[ \cross{\bsig}{\mathbf{p}_0}{} A_0 \right] \cdot \mathbf{A} \psi \, , \label{eq:L_QED_NR}
\end{align}
where $P_+ = (1 + \gamma^0)/2$ is only non-zero in the upper left diagonal component, the trace is in $2 \times 2 $ block diagonal space, and we have ignored the $\psi_h$ terms. The Pauli matrices are given by $\bm{\sigma}$, the momentum variables ($\mathbf{k}, \mathbf{p}, \omega'$) are shorthand for derivatives, $p^\mu = i \partial^\mu$, $\mathbf{k}$ acts on $\psi$, $\mathbf{p}$ and $\omega'$ act on $\mathbf{A}$, $\mathbf{p}_0$ acts on $A_0$, and $\mathbf{K}' = 2 \mathbf{k} + \mathbf{p}$.

\subsection{NR Limit of MDM and EDM Interactions}
\label{subapp:op_UV_to_NR}

With the field redefinition which diagonalizes QED Lagrangian found in Eq.~\eqref{eq:FW_field_redef}, we can now use this to expand the interactions in Eqs.~\eqref{eq:L_mdm} and \eqref{eq:L_edm}. However before this we can reach a more general result: the NR limit of a UV general operator, to any order in $1/m_e$ (still only leading order in DM-electron coupling),
\begin{align}
    \bar{\Psi} \mathcal{O} \Psi \rightarrow \psi^\dagger \left[ \exp\left( i \frac{X_{n - 1}}{m_e^n} \right) \ldots \exp\left( i \frac{X_{0}}{m_e} \right) \gamma^0 \mathcal{O} \exp\left( -i \frac{X_{0}}{m_e} \right) \ldots \exp\left( -i \frac{X_{n - 1}}{m_e^n} \right) \right] \psi \, ,
    \label{eq:general_op_expansion}
\end{align}
where the outer most brackets, $\left[ \cdots \right]$, implicitly indicate taking the $2 \times 2$ upper diagonal component of the $4 \times 4$ matrix. All $X$ act to the right, meaning that those on the left side of $\mathcal{O}$ will also act on $\mathcal{O}$. To $\mathcal{O}(m_e^{-2})$ this expression simplifies to,
\begin{align}
    \bar{\Psi} \mathcal{O} \Psi & \approx \psi^\dagger \left[ \exp\left( i \frac{X_{1}}{m_e^2} \right) \left( \gamma^0 \mathcal{O} + \frac{i}{m_e} \left[ X_0, \gamma^0 \mathcal{O} \right] - \frac{1}{m_e^2} \left[ (X_0)^2, \gamma^0 \mathcal{O} \right]  \right) \exp\left( -i \frac{X_1}{m_e^2} \right) \right] \psi \\
    \approx & \; \psi^\dagger \left[ \gamma^0 \mathcal{O} + \frac{i}{m_e} \left[ X_0, \gamma^0 \mathcal{O} \right] - \frac{1}{m_e^2} \left[ (X_0)^2, \gamma^0 \mathcal{O} \right] + \frac{i}{m_e^2} \left[ X_1, \gamma^0 \mathcal{O} \right]  \right] \psi \, , \label{eq:expanded_general_op}
\end{align}
where $\left[ (A)^2, B \right] \equiv [A, [A, B]]$ and we have made use of the Campbell-Baker-Hausdorff formula when expanding matrix
products of the form $e^X Y e^{-X}$.

Focusing on Eqs.~\eqref{eq:L_mdm} and \eqref{eq:L_edm}, in three component notation we can expand the operators as,
\begin{align}
    \frac{d_M}{2} V_{\mu \nu} \bar{\Psi} \sigma^{\mu \nu} \Psi & = d_M ( -\partial^0 V^i + \partial^i V^0 ) \bar{\Psi} \sigma^{0i} \Psi + d_M \partial^i V^j \bar{\Psi} \sigma^{ij} \Psi \label{eq:expanded_MDM}\\
    \frac{i d_E}{2} V_{\mu \nu} \bar{\Psi} \sigma^{\mu \nu} \gamma^5 \Psi & = i d_E ( -\partial^0 V^i + \partial^i V^0 ) \bar{\Psi} \sigma^{0i} \gamma^5 \Psi + i d_E \partial^i V^j \bar{\Psi} \sigma^{ij} \gamma^5 \Psi \, , \label{eq:expanded_EDM}
\end{align}
and therefore we see that there are four $\mathcal{O}$ operators whose NR limit must be extracted:
\be
\bar{\Psi} \sigma^{0i} V \Psi~~, ~~\bar{\Psi} \sigma^{ij} V \Psi ~~,~~ \bar{\Psi} \sigma^{0i} \gamma^5 V \Psi ~~,~~ \bar{\Psi} \sigma^{ij} \gamma^5 V \Psi \,.  
\ee
Each of the terms in Eq.~\eqref{eq:expanded_general_op} can be further simplified, defining $\Gamma \equiv \gamma^0 \mathcal{O}$,
\begin{align}
    \left[ X_0, \Gamma V \right] & = \frac{1}{2} \left( \gamma_i \Gamma \left( \partial_i V \right) + \left[ \gamma_i, \Gamma \right] V D_i \right) \\ 
    \left[ X_1, \Gamma V \right] & = -\frac{e}{4} \left[ \gamma_0 \gamma_i, \Gamma \right] V F_{0i} \\
    \left[ (X_0)^2, \Gamma V \right] & = \left[ X_0, \left[ X_0, \Gamma V \right] \right] \nonumber \\
    & = \frac{1}{4} \left( \left[ \gamma_i, \left[ \gamma_j , \Gamma \right] \right] \right) V D_i D_j + \frac{1}{4} \gamma_i \left[ \gamma_j, \Gamma \right] (\partial_i V) D_j + \frac{1}{4} \left[ \gamma_i, \gamma_j \Gamma \right] (\partial_j V) D_i + \frac{1}{4} \gamma_i \gamma_j \Gamma (\partial_i \partial_j V) \nonumber \\
    &\quad+ \frac{ie}{4} \left[ \gamma^j, \Gamma \right] \gamma^i V F_{ij} \, ,
\end{align}
where $\left( \partial_i V \right)$ indicates that the derivative operator only acts on $V$. The final NR expansion, to $\mathcal{O}(m_e^{-2})$, is given in Table~\ref{tab:nr_limit}.

\renewcommand{\arraystretch}{2}
\begin{table}
    \centering
    \begin{tabularx}{\textwidth}{c|c}
        \toprule
            $\displaystyle \mathcal{O}_\text{UV}$ & $\displaystyle \mathcal{O}_\text{NR}$  \\[4pt]
        \midrule \\
            \;$\displaystyle \sigma^{0 i} V$ \;& $\begin{aligned} & \frac{1}{2m_e} \left( \dV{i} + i \epsilon^{ijk} \sigma^j \left( \dV{k} + 2 V \partial^k \right) \right) - \frac{e V}{m_e} \left( \epsilon^{ijk} \sigma^j A^k \right) - \frac{eV}{2m_e^2} \left( \left[ \partial^0 A^i \right] - \left[ \partial^i A^0 \right] \right) \end{aligned}$ \\[40pt]
            \;$\displaystyle \sigma^{i j} V$ \;& \;$\begin{aligned} & \epsilon^{ijk} \left( V\sigma^k + \frac{1}{2m_e^2} \left( \dV{k} \sigma^m \partial^m + \sigma^m \left( 2 V \partial^m + \dV{m} \right) \partial^k - i \epsilon^{kmn} \dV{m} \partial^n + \frac{1}{2} \left[ \partial^2 V \right] \sigma^k \right) \right) \\
        & + \frac{i e\epsilon^{ijk}}{2m_e^2} \left( \sigma^m A^m \left( 2 V\partial^k + \dV{k} \right) + A^k \sigma^m \left( 2V\partial^m + \dV{m} \right) - V \sigma^m \left[ \partial^k A^m \right] + 3 V \sigma^m \left[ \partial^m A^k \right] + i \epsilon^{kmn} A^m \dV{n} \right)\end{aligned}$ \\[40pt]
            \;$\displaystyle i\sigma^{0 i} \gamma^5 V$\; &  \;$\begin{aligned} & -\sigma^i V - \frac{1}{2m_e^2} \left( \dV{i} \sigma^m \partial^m + \sigma^m \left( 2 V \partial^m + \dV{m} \right) \partial^i - i \epsilon^{imn} \dV{m} \partial^n + \frac{1}{2} \left[ \partial^2 V \right] \sigma^i \right) \\ & - \frac{ie}{2m_e^2} \left( \sigma^m A^m \left( 2V \partial^i + \dV{i} \right) + A^i \sigma^m \left( 2V\partial^m + \dV{m} \right) - V \sigma^m \left[ \partial^i A^m \right] + 3 V \sigma^m \left[ \partial^m A^i \right] + i \epsilon^{imn} A^m \dV{n}  \right) \end{aligned}$ \\[40pt]
            \;$i\sigma^{i j} \gamma^5 V$\; & \;$\begin{aligned} -\frac{\epsilon^{ijk}}{2m_e} \left( \dV{k} + i \epsilon^{kmn} \sigma^m \left( 2V \partial^n + \dV{n} \right) \right) - \frac{e V \epsilon^{ijk}}{m_e} \epsilon^{kmn} \sigma^m A^n - \frac{eV \epsilon^{ijk}}{2m_e^2} \left( \left[ \partial^0 A^k \right] - \left[ \partial^k A^0 \right] \right) \end{aligned}$ \\[20pt]
        \bottomrule
    \end{tabularx}
    \caption{NR limit (right column), $\bar{\Psi} \mathcal{O}_\text{UV} \Psi \rightarrow \psi^\dagger \mathcal{O}_\text{NR} \psi$, of the relevant UV operators (left column) needed to find the NR limit of the MDM and EDM interactions in Eqs.~\eqref{eq:L_mdm} and \eqref{eq:L_edm}, respectively. We only keep terms involving two fields (excluding the electron field, $\psi$).}
    \label{tab:nr_limit}
\end{table}
We can now substitute the results in Table~\ref{tab:nr_limit} to the terms in Eqs.~\eqref{eq:expanded_MDM} and \eqref{eq:expanded_EDM} to find the NR limit of the Lagrangians given in Eqs.~\eqref{eq:L_mdm} and \eqref{eq:L_edm},
\begin{align}
     \mathcal{L}_\text{MDM}^\text{NR}& \approx \frac{d_M}{m_e} \psi^\dagger\left[ -\frac{\mathbf{q}^2}{2} - i \mathbf{q} \cdot \left( \bm{\sigma} \times \mathbf{k} \right)  \right] V_0 \psi \nonumber\\
     & + d_M \psi \left[ -i (\bm{\sigma} \times \mathbf{q}) \left( 1 - \frac{q^2}{4m_e^2} \right) + \frac{\omega}{2m_e} \left( \mathbf{q} + i \cross{\bsig}{\mathbf{K}}{} \right) + \frac{1}{2m_e^2} \left( -i \cross{\mathbf{q}}{\mathbf{k}}{} \left( \bm{\sigma} \cdot \mathbf{K} \right) + q^2 \mathbf{k} - \left( \mathbf{q} \cdot \mathbf{k} \right) \mathbf{q} \right) \right] \cdot \mathbf{V} \psi \nonumber \\
     & - \frac{e d_M}{2m_e^2} \psi^\dagger \left[ \left( \mathbf{p} \cdot \mathbf{q} \right) \right] A_0 V_0 \psi \nonumber \\
     & + e d_M \psi^\dagger \left[ \frac{i}{m_e} \left( \bm{\sigma} \times \mathbf{q} \right) + \frac{\omega'}{2m_e^2} \mathbf{q} \right] \cdot \mathbf{A} V_0 \psi \nonumber \\
    & + e d_M \psi^\dagger \bigg[ -\frac{\omega}{m_e} i \cross{\bsig}{\mathbf{A}}{} \nonumber \\
    & \quad\quad\quad\quad\quad + \frac{1}{2m_e^2} \left( i\left( \bm{\sigma} \cdot \left( \mathbf{K} + 3 \mathbf{p} \right) \right) \left( \mathbf{q} \times \mathbf{A} \right) + i \left( \bm{\sigma} \cdot \mathbf{A} \right) \left( \mathbf{q} \times \left( \mathbf{k} - \mathbf{p} \right) \right) - q^2 \mathbf{A} + \left( \mathbf{q} \cdot \mathbf{A} \right) \mathbf{q} - \omega \omega' \mathbf{A} \right) \bigg] \cdot \mathbf{V} \psi \nonumber \\
        & + \frac{e d_M \omega }{2m_e^2} \psi^\dagger \left[ \mathbf{p} \right] \cdot \mathbf{V} A_0 \psi \label{eq:L_NR_MDM_app}
\end{align}
\begin{align}
    \mathcal{L}_\text{EDM}^\text{NR} & \approx i d_E \psi^\dagger \left[ \mathbf{q} \cdot \bm{\sigma} \left( 1 - \frac{\mathbf{q}^2}{4m_e^2} \right) + \frac{1}{2m_e^2} \left(- \left( \mathbf{k} \cdot \mathbf{q} \right) \left( \bm{\sigma} \cdot \mathbf{K} \right) - q^2 \left( \bm{\sigma} \cdot \mathbf{k} \right) \right) \right] V_0 \psi \nonumber\\
    & + i d_E \psi^\dagger \left[ -\omega \bm{\sigma} \left( 1 - \frac{q^2}{4m_e^2} \right) + \frac{1}{2m_e} \left( -\left( \bm{\sigma} \cdot \mathbf{q} \right) \mathbf{K} + \left( \mathbf{q} \cdot \mathbf{K} \right) \bm{\sigma} \right) + \frac{\omega}{2m_e^2} \left( \left( \bm{\sigma} \cdot \mathbf{K} \right)\mathbf{k} + \left( \bm{\sigma} \cdot \mathbf{k} \right) \mathbf{q} - i \cross{\mathbf{q}}{\mathbf{k}}{} \right) \right] \cdot \mathbf{V} \psi \nonumber \\
    & + \frac{i e d_E}{2m_e^2} \psi^\dagger \left[ \left( \mathbf{q} \cdot \mathbf{K} - \mathbf{p} \cdot \mathbf{q} \right) \bm{\sigma} + \left( \bm{\sigma} \cdot \mathbf{K} + 3 \left( \bm{\sigma} \cdot \mathbf{p} \right) \right) \mathbf{q}  \right] \cdot \mathbf{A} V_0 \psi \nonumber \\
    & + i e d_E \psi^\dagger \bigg[ \frac{1}{m_e} \left( \left( \bm{\sigma} \cdot \mathbf{q} \right) \mathbf{A} - \left( \mathbf{q} \cdot \mathbf{A} \right) \bm{\sigma} \right) - \frac{\omega'}{2m_e^2} \left( \mathbf{q} \times \mathbf{A} \right) \nonumber\\
    &\quad\quad\quad\quad\quad + \frac{\omega}{2 m_e^2} \left( - (\bm{\sigma} \cdot \mathbf{A}) \left( \mathbf{K} - \mathbf{p} \right) - \left( \bm{\sigma} \cdot \left( \mathbf{K} + 3 \mathbf{p} \right) \right) \mathbf{A} - \left( \mathbf{q} \times \mathbf{A} \right) \right)  \bigg] \cdot \mathbf{V} \psi \nonumber \\
    & - \frac{ed_E}{2m_e^2} \psi^\dagger \left[ \mathbf{p} \times \mathbf{q} \right] \cdot \mathbf{V} A_0 \psi \, , \label{eq:L_NR_EDM_app}
\end{align}
where, similar to Sec.~\ref{subapp:nrqed_FW}, the momentum are shorthand for derivatives acting on different fields and $\omega, \mathbf{q}$ act on $V$, $\mathbf{p}, \omega'$ act on $A$, and $\mathbf{k}$ acts on $\psi$.  

\section{Self-Energy Calculations}
\label{app:self_energy_details}

With the NR limit of the QED, MDM, and EDM Lagrangians, given in Eqs.~\eqref{eq:L_QED_NR}, \eqref{eq:L_NR_MDM_app} and \eqref{eq:L_NR_EDM_app}, respectively, we can now derive the self-energies needed to compute the rate in Eq.~\eqref{eq:abs_rate_self_energies}. To do this in full generality, one must first compute the diagonal self-energies in the component basis, i.e., $\Pi^{\mu \nu}_{VV}$ and $\Pi^{\mu \nu}_{AA}$, then find the polarization vectors, $\epsilon^\mu_\lambda$, such that they are diagaonlized, i.e.,
\begin{align}
    \Pi^{\mu \nu}_{\phi \phi} = - \sum_{\lambda} \epsilon^\mu_\lambda \, \Pi_{\phi \phi}^\lambda \, \epsilon^{\nu, *}_\lambda \, . \label{eq:self_energy_projection}
\end{align}
where $\phi \in \{ V, A \}$. The off-diagonal self-energies, $\Pi_{VA}^{\lambda \lambda'}$ are then simply the off-diagonal self-energies in the component basis, $\Pi_{VA}^{\mu \nu}$, projected on to the polarization vectors.

For the targets of interest (Xe, Ar, Si, and Ge) we can make a key simplifying assumption: isotropy. Assuming that the targets are isotropic the spatial components of the self-energies become,
\begin{align}
    \Pi^{ij} = \frac{1}{3} \delta^{ij} \Pi^{kk} \, ,
\end{align}
where the repeated $k$ index on the right hand side is summed. Under this assumption it can be shown that the polarizations which diagonalize $\Pi_{\phi \phi}^{\mu \nu}$ are given by the standard longitudinal and transverse polarization vectors,
\be
    \epsilon^\mu_\pm  = (0, \hat{\mathbf{q}}_\pm) ~~~,~~~
    \epsilon^\mu_L = \frac{1}{\sqrt{Q^2}}(q, \omega \hat{\mathbf{q}})
\ee
where $\mathbf{q}_\pm$ are two orthogonal vectors satisfying $\mathbf{q} \cdot \mathbf{q}_\pm = 0$ and $Q^2 = \omega^2 - q^2$. 

The Ward Identity (WI) simplifies the calculation of projecting the component basis self-energies to the polarization basis. One can show that,
\begin{align}
    \epsilon_L^\mu \Pi_{\mu \nu} \epsilon_L^\nu = \frac{Q^2}{q^2} \Pi^{00} \, ,
\end{align}
and therefore the longitudinal self-energy can be computed from $\Pi^{00}$ alone. Additionally, the transverse components are only related to $\Pi^{ij}$. Therefore these are the only self-energies we need to compute.

In Secs.~\ref{subapp:self_energy_qed}, \ref{subapp:self_energy_mdm} and \ref{subapp:self_energy_edm} we derive the leading order contributions to the self-energies in the component basis, and then project them on to the polarization basis, i.e., the $T, L$ components, via the inverse of Eq.~\eqref{eq:self_energy_projection}. Finding the leading order contribution is a relatively straightforward exercise in power counting with respect to the dimensionless variables, $v_e = k/m_e$, $v_V = q/m_V$ and $m_V / m_e$, where $v_e$ is the electron velocity, $q$ is the dark photon momentum, and $m_V$ is the dark photon mass. The main subtlety, discussed in detail in Ref.~\cite{Mitridate:2021ctr}, is due to state orthonormality which reduces the order of operators. For example, while an interaction of the form, $A_0 \psi^\dagger \psi$ is naively $\mathcal{O}(1)$, the reduced self-energy, $\bPi$, Eq.~\eqref{eq:reduced_self_energy}, depends on
\be
\langle F | e^{i \mathbf{q} \cdot \mathbf{x} } | I \rangle \approx i \mathbf{q} \cdot \langle F | \mathbf{x} | I \rangle \sim \mathcal{O}( v_e v_V )~,
\ee
and therefore this is a suppressed operator. 

At the end of each section we find that $\Pi_T \approx \Pi_L$ (these are the self-energies projected on to the transverse and longitudinal polarization vectors), and therefore $\Pi_\lambda$ is approximately independent of $\lambda$, justifying the simplification in Sec.~\ref{sec:absorption_calculation}. Additionally, we find that the mixing between $V, A$ in the EDM model is negligible, and therefore the section deriving this is absent.

The self-energies resulting from these calculations will be written in terms of some ``reduced" self-energies $\bPi_{\mathcal{O}_1, \mathcal{O}_2}$, which are only dependent on the target electronic structure. Ref.~\cite{Mitridate:2021ctr, Trickle:2022fwt} details how these are computed for crystal targets, and in Sec.~\ref{subapp:atomic_target_self_energy} we derive the formula for atomic targets.

\subsection{QED}
\label{subapp:self_energy_qed}

Starting from the NR limit of the QED Lagrangian in Eq.~\eqref{eq:L_QED_NR}, one can derive the photon self-energies,
\begin{align}
    \Pi_{AA}^{00}  = -e^2 \bPi_{1, 1} ~~~~,~~~~
    \Pi_{AA}^{ij}  = -\frac{e^2}{m_e^2} \bPi_{k^i, k^j} + \frac{e^2}{m_e} \delta^{ij} \bPi_{1}
\end{align}
where the single $\mathcal{O}$ reduced self-energy, $\bPi_\mathcal{O}$, is given by,
\begin{align}
    \bPi_\mathcal{O} \equiv - \frac{1}{V} \sum_I \langle I | \mathcal{O} | I \rangle \, ,
\end{align}
where $I$ runs over filled electronic states. Note that this result was also derived in Ref.~\cite{Mitridate:2021ctr}. The $\bPi_1$ term can be reduced using a WI, $\omega \Pi^{0 \mu} - q^i \Pi^{i\mu} = 0$, to,
\begin{align}
    \Pi_{AA}^{00}  = -e^2 \bPi_{1, 1} ~~~,~~~
    \Pi_{AA}^{ij}  = \frac{e^2}{q^2} \delta^{ij} \left[ \frac{1}{m_e^2} q^i \bPi_{k^i, k^j} q^j - \omega^2 \bPi_{1, 1} \right] - \frac{e^2}{m_e^2} \bPi_{k^i, k^j} \, .
\end{align}
Under our isotropic target approximation we can replace $\bPi_{k^i, k^j} \rightarrow \frac{1}{3} \bPi_{k^i, k^i} \delta^{ij}$, thereby reducing the self-energies to,
\begin{align}
    \Pi_{AA}^{00} = -e^2 \bPi_{1, 1} ~~~,~~~
    \Pi_{AA}^{ij}  = -\frac{ e^2 \omega^2}{q^2}  \bPi_{1, 1} \delta^{ij} \, .
\end{align}
Lastly, projecting on to the polarizations gives,
\begin{align}
    \Pi^{\pm}_{AA}  = \frac{e^2 \omega^2}{q^2} \bPi_{1, 1} ~~~,~~~
    \Pi^L_{AA}  = \frac{e^2 Q^2}{q^2} \bPi_{1, 1} \approx \frac{e^2 \omega^2}{q^2} \bPi_{1, 1} \, .
\end{align}
Note that explicit expressions for $\bar \Pi_{{\cal O}_1, {\cal O }_2}$  are computed 
in Refs.~\cite{Chen:2022pyd, Mitridate:2021ctr, Trickle:2022fwt} for crystal targets and
in Appendix \ref{app:atomic_transition_matrix_elements} below for atomic targets. 

\subsection{Magnetic Dipole Moment}
\label{subapp:self_energy_mdm}

For the MDM model two types of self-energies need to be computed: the dark photon self-energy with itself, $\Pi_{VV}$, and the mixing of the dark photon with the photon via $\Pi_{VA}$. For readability we split these two calculations in to the subsections below.

\subsubsection{Dark Photon Self-Energy}

Starting from the MDM interaction in Eq.~\eqref{eq:L_NR_MDM_app} the leading order contribution to the self-energies in the component basis are,
\begin{align}
    \Pi_{VV}^{00}  = - \frac{d_M^2}{m_e^2} \bPi_{ \mathbf{q} \cdot \cross{\bsig}{\mathbf{k}}{}, \mathbf{q} \cdot \cross{\bsig}{\mathbf{k}}{} }
    ~~,~~
    \Pi_{VV}^{ij}  = -\frac{d_M^2 \omega^2}{m_e^2} \bPi_{ \cross{\bsig}{\mathbf{k}}{i}, \cross{\bsig}{\mathbf{k}}{j} } \label{eq:no_spin_order_ij}\, .
\end{align}
The dominant term in Eq.~\eqref{eq:L_NR_MDM_app}, when computing, e.g., $\Pi_{VV}^{ij}$, can be easily extracted. Simply take the $\mathbf{q} \rightarrow 0$ limit of terms involving $\mathbf{V} \psi^\dagger \psi$ and there is only a single remaining term. To further reduce these expressions we must make an additional approximation relative to the photon self-energy calculations. We assume that the target is not spin-ordered, and therefore the sums over initial and final states can be split in to $\sum_I \rightarrow \sum_i \sum_s$, where $s$ indexes the spin states. This allows the Pauli matrices to be traced over in the reduced self-energy expressions, e.g., in Eq.~\eqref{eq:no_spin_order_ij}, 
\begin{align}
    \bPi_{ (\bm{\sigma} \times \mathbf{k})^i, (\bm{\sigma} \times \mathbf{k})^j} \rightarrow \delta^{ij} \bPi_{k^i, k^i} - \bPi_{k^i, k^j} \, .
\end{align}
Note that our convention for non spin-ordered targets is to absorb the factor of two, from the Pauli spin matrix trace, in to $\bPi$. With this substitution the equations in Eq.~\eqref{eq:no_spin_order_ij} become,
\begin{align}
    \Pi_{VV}^{00}  = -\frac{d_M^2}{m_e^2} \left( q^2 \bPi_{k^i, k^i} - q^i q^j \bPi_{k^i, k^j} \right) ~~~,~~~
    \Pi_{VV}^{ij}  = -\frac{d_M^2 \omega^2}{m_e^2} \left( \delta^{ij} \bPi_{k^i, k^i} - \bPi_{k^i, k^j} \right) \, .
\end{align}
Using the isotropic approximation, analogous to Sec.~\ref{subapp:self_energy_qed}, these self-energies simplify to,
\begin{align}
    \Pi_{VV}^{00}  = -\frac{2 d_M^2 q^2}{3 m_e^2} \bPi_{k^i, k^i} ~~~,~~~
    \Pi_{VV}^{ij}  = -\frac{2 d_M^2 \omega^2}{3m_e^2} \delta^{ij} \bPi_{k^i, k^i} \, .
\end{align}
Lastly, we project these into the polarization basis to reach our final results,
\begin{align}
    \Pi_{VV}^\pm  = -\frac{2 d_M^2 \omega^2}{3m_e^2} \bPi_{k^i, k^i} ~~~,~~~
    \Pi_{VV}^L  = -\frac{2 d_M^2 \omega^2 }{3 m_e^2} \bPi_{k^i, k^i} \, .
\end{align}

\subsubsection{Dark Photon-Photon Mixed Self-Energy}

The dark photon-photon mixed self-energy is computed from the MDM and QED Lagrangians in Eqs.~\eqref{eq:L_NR_MDM_app} and \eqref{eq:L_NR_EDM_app}, respectively. Note that while we will only compute $\Pi_{VA}$, at this order $\Pi_{VA} = \Pi_{AV}$. The leading order mixed self-energies are given by,
\begin{align}
    \Pi_{VA}^{00}  = - e d_M \frac{q^2}{2m_e^2} \bPi_{1} + \frac{e d_M}{4 m_e^3} q^i q^j \bPi_{\cross{\bsig}{\mathbf{k}}{i}, \cross{\bsig}{\mathbf{k}}{j} } 
    ~~~,~~~
    \Pi_{VA}^{ij}  = - e d_M \frac{\omega^2}{2m_e^2} \delta^{ij} \bPi_{1} + \frac{e d_M}{4 m_e^3} \omega^2 \bPi_{\cross{\bsig}{\mathbf{k}}{i}, \cross{\bsig}{\mathbf{k}}{j} } \, ,
\end{align}
where, similar to the previous section, we have assumed that the target is not spin ordered, i.e., $\bPi_{\bm{\sigma}} = 0$. This assumption also allows us to trace out the $\sigma$ matrices,
\begin{align}
    \Pi_{VA}^{00}  = -\frac{e d_M q^2}{2m_e^2} \left( \bPi_{1} - \frac{1}{3 m_e} \bPi_{k^i, k^i} \right) 
    ~~~,~~~
    \Pi_{VA}^{ij}  = - \frac{e d_M \omega^2}{2m_e^2} \delta^{ij} \left( \bPi_{1} - \frac{1}{3m_e} \bPi_{k^i, k^i} \right) \, .
\end{align}
As done in Sec.~\ref{subapp:self_energy_qed} we now replace $\bPi_1$ via a WI, and utilize our isotropic assumption to get to,
\begin{align}
    \Pi_{VA}^{00}  = \frac{ed_M \omega^2}{2m_e} \bPi_{1, 1} ~~~,~~~
    \Pi_{VA}^{ij}  = \frac{ed_M \omega^2}{2m_e} \frac{\omega^2}{q^2} \delta^{ij} \bPi_{1, 1} \, .
\end{align}
Lastly, projecting on to components gives,
\begin{align}
    \Pi_{VA}^\pm  = -\frac{ed_M \omega^4}{2m_e q^2} \bPi_{1, 1} ~~~,~~~
    \Pi_{VA}^L  = -\frac{e d_M \omega^4}{2m_e q^2} \bPi_{1, 1} \, ,
\end{align}
which are related to the photon self-energies given in Sec.~\ref{subapp:self_energy_qed}.

\subsection{Electric Dipole Moment}
\label{subapp:self_energy_edm}

The dark photon self-energy in the EDM model is derived from the Lagrangian in Eq.~\eqref{eq:L_NR_EDM_app}. In the component basis they are given by,
\begin{align}
    \Pi_{VV}^{00}  = -\frac{d_E^2}{m_e^4} q^i q^j \bPi_{k^i (\sigma \cdot \mathbf{k}), k^j (\sigma \cdot \mathbf{k}) }
    ~~~,~~~
    \Pi_{VV}^{ij}  = - \frac{d_E^2 \omega^2}{m_e^4} \bPi_{ k^i (\sigma \cdot \mathbf{k}), k^j (\sigma \cdot \mathbf{k})} \, .
\end{align}
Again assuming no spin ordering and target isotropy we can replace,
\begin{align}
    \bPi_{ k^i (\sigma \cdot \mathbf{k}), k^j (\sigma \cdot \mathbf{k})} \rightarrow \bPi_{k^i k^\alpha, k^j k^\alpha} \rightarrow \frac{1}{3} \delta^{ij} \bPi_{k^i k^j, k^i k^j} \, ,
\end{align}
leading to,
\begin{align}
    \Pi_{VV}^{00}  = - \frac{d_E^2 q^2}{3m_e^4}\bPi_{k^i k^j, k^i k^j }~~~,~~~
    \Pi_{VV}^{ij}  = - \frac{d_E^2 \omega^2}{3m_e^4} \bPi_{ k^i k^j, k^i k^j} \, .
\end{align}
Lastly, projecting on to components gives,
\begin{align}
    \Pi_{VV}^{\pm}  = \frac{d_E^2 \omega^2}{3m_e^4} \bPi_{k^i k^j, k^i k^j }~~~,~~~
    \Pi_{VV}^{L}  = \frac{d_E^2 \omega^2}{3m_e^4} \bPi_{ k^i k^j, k^i k^j} \, .
\end{align}

\subsection{Atomic Target Self-Energy}
\label{subapp:atomic_target_self_energy}

In this subsection, we simplify the reduced self-energy, also given in Eq.~\eqref{eq:reduced_self_energy},
\begin{align}
    \bar{\Pi}_{\mathcal{O}_1, \mathcal{O}_2} & = \frac{1}{V} \sum_{IF} \frac{G(\omega, \omega_F - \omega_I, \delta)}{\langle I | I \rangle \langle F | F \rangle} \mathcal{T}_{\mathcal{O}_1} \mathcal{T}_{\mathcal{O}_2}^*\, , \label{eq:reduced_self_energy_app}
\end{align}
for the electronic states in an atomic target. More detailed specifics about the electronic states are discussed in Sec.~\ref{app:atomic_transition_matrix_elements}. While our focus will be on an atomic target with a single atomic species, the analysis here easily generalizes to multiple atomic species. In such a target there are $N = n \times V$ target atoms, where $n$ is the number density and $V$ is the target volume. Each atom hosts bound electronic states which are indexed by the quantum numbers $n, \ell, m$. Assuming these states are spin degenerate, the sum over initial electron states in Eq.~\eqref{eq:reduced_self_energy_app} becomes,
\begin{align}
    \sum_I & \rightarrow 2 \times n \times V \times \sum_{n \ell m} \, .
\end{align}

The discrete nature of this sum makes this substitution intuitive. On the other hand, the final states form a continuum indexed by $k, \ell', m'$ where $\ell', m'$ are angular quantum numbers, and $k$ is a continuous momentum. The sum over $k$ then becomes an integral,
\begin{align}
    \sum_F & \rightarrow \sum_{k\ell'm'} \rightarrow \delta(0) \int dk \sum_{\ell' m'} \, ,
\end{align}
where $\delta(0)$ is a normalization coefficient which will drop out of the final formulas. The simplest way to take care of this is to define the initial (discrete, bound) and final (continuum) states with discrete and continuous normalizations,
\begin{align}
    \langle I | I \rangle & = 1 \\
    \langle F | F \rangle & = 2 \pi \delta(0) \, .
\end{align}
Substituting these sums and state normalizations leads to a reduced self-energy of,
\begin{align}
    \bar{\Pi}_{\mathcal{O}_1, \mathcal{O}_2} = 2 n \sum_{n \ell \ell' m m'} \int \frac{dk}{2 \pi} G(\omega, \omega_F - \omega_I, \delta) \mathcal{T}_{\mathcal{O}_1} \mathcal{T}_{\mathcal{O}_2}^* \, ,
\end{align}
where $\delta$ is the electron width. Further simplifications can be made to isolate the imaginary part of this self-energy needed to compute the rate in Eqs.~\eqref{eq:rate_MDM} and \eqref{eq:rate_EDM}. In the limit of zero electron width, $\delta \rightarrow 0$, the Green's function, $G$, simplifies,
\begin{align}
    \lim_{\delta \rightarrow 0} \text{Im} \left[ G(\omega, \Delta \omega, \delta) \right] = - \pi \delta(\omega - \Delta \omega) \, .
\end{align}
This delta function reduces the $k$ integral, since $\omega_F = k^2/2m_e$, such that,
\begin{align}
    \lim_{\delta \rightarrow 0} \text{Im} \left[ \bar{\Pi}_{\mathcal{O}_1, \mathcal{O}_2} \right]& = -2n \sum_{n\ell\ell'mm'} \int \frac{dk}{2} \, \delta \left( \frac{k^2}{2m_e} - \omega - \omega_I \right) \mathcal{T}_{\mathcal{O}_1}\mathcal{T}_{\mathcal{O}_2}^* \\
    & =  -n m_e \sum_{n\ell\ell'mm'} \frac{1}{\sqrt{2 m_e (\omega + \omega_I)}} \mathcal{T}_{\mathcal{O}_1}\mathcal{T}_{\mathcal{O}_2}^* \, ,
\end{align}
and the imaginary part of the reduced self-energy becomes a simple sum. The transition matrix elements, $\mathcal{T}_\mathcal{O}$, are derived in App.~\ref{app:atomic_transition_matrix_elements}.

\section{Atomic Absorption Transition Matrix Elements}
\label{app:atomic_transition_matrix_elements}

To compute the self-energies derived in Sec.~\ref{app:self_energy_details}, the transition matrix elements, $\mathcal{T}_\mathcal{O} \equiv \langle F | \mathcal{O} | I \rangle$ must be computed. The details of this calculation for crystal targets, e.g., Si and Ge, has been discussed in Refs.~\cite{Mitridate:2021ctr, Trickle:2022fwt}, and similar calculations for transitions between atomic bound and continuum states have been performed in the scattering limit~\cite{Catena:2019gfa, Catena:2022fnk, Essig:2015cda}. In this appendix we derive the transition matrix elements between the bound and continuum electronic states in an isolated atom (appropriate for Xe and Ar targets), in the absorption limit.

We begin by defining the initial, $| I \rangle$, and final, $| F \rangle$, quantum states (mainly for comparison with other conventions). The initial, bound states are labeled by the standard quantum numbers, $n, \ell, m$, and satisfy,
\begin{align}
    \langle n' \ell' m' | n \ell m \rangle = \delta_{nn'} \delta_{\ell \ell'} \delta_{mm'} \, .
    \label{eq:ortho_initial}
\end{align}
Note that the states are dimensionless. The position space representation of these states is,
\begin{align}
    \langle \mathbf{x} | n \ell m \rangle  \equiv \sqrt{V} \psi_{n\ell m}(\mathbf{x}) = \sqrt{V} R_{n \ell}(x) Y_{\ell m}(\hat{\mathbf{x}}) \, ,
\end{align}
where $\langle \mathbf{x} | \mathbf{x} \rangle = 1$, and $\psi_{n\ell m}$ has dimension $\mathrm{eV}^{3/2}$. We assume that the position space representation of these states can be further expanded as,
\begin{align}
    \psi_{n\ell m}(\mathbf{x}) = R_{n \ell}(x) Y_{\ell m}(\hat{\mathbf{x}}) \, ,
\end{align}
where $R_{n \ell}$ has dimension $\mathrm{eV}^{3/2}$, and $Y_{\ell m}$ are the spherical harmonics, normalized to,
\begin{align}
    \int d \Omega \,  Y_{\ell' m'}^* Y_{\ell m} = \delta_{\ell \ell'} \delta_{m m'} \, .
\end{align}
In order to satisfy the orthonormality relationship in Eq.~\eqref{eq:ortho_initial}, the $R_{n \ell}$ must satisfy,
\begin{align}
    \int dr \, r^2 R_{n' \ell}^* R_{n \ell} = \delta_{n n'} \, .
\end{align}
A useful basis to expand $R_{n \ell}$ in is the ``Slater type orbital" basis,
\begin{align}
    R_{n \ell}(r) & = \sum_{j} C_{n \ell j} R_\text{STO}(r; Z_{\ell j}, n_{\ell j}) \\
    R_\text{STO}(r; Z, n)& = a_0^{- 3/2} \frac{(2Z)^{n + \frac{1}{2}}}{\sqrt{(2n)!}} \left( \frac{r}{a_0} \right)^{n - 1} e^{-Z r / a_0}
\end{align}
where $a_0$ is the Bohr radius, $C_{n, \ell, j}, n_{\ell, j}, Z_{\ell j}$ are coefficients found by solving the isolated atom Hamiltonian. We use the coefficients tabulated in Ref.~\cite{Bunge_1993} to compute the results in the main text.

The final states are taken to be the Coloumb wave function solutions to a $-Z/r$ potential~\cite{DarkSide:2018ppu,Catena:2019gfa,Catena:2022fnk,Peng_2010,Sabbatucci_2016,Tan:2021nif}. These states are labelled by $k, \ell, m$, where $k$ is a continuous index. Different conventions are reasonable for orthonormalizing these states; here we choose,
\begin{align}
     \langle k' \ell' m' | k \ell m \rangle = 2 \pi \delta(k - k') \delta_{\ell \ell'} \delta_{m m'}\, .
     \label{eq:ortho_final}
\end{align}
The position space representation can be decomposed in a way analogous to the initial states,
\begin{align}
    \langle \mathbf{x} | k \ell m \rangle = \sqrt{V} \psi_{k \ell m}(\mathbf{x}) = \sqrt{V} R_{k \ell}(x) Y_{\ell m}(\hat{\mathbf{x}}) \, , \label{eq:atomic_state_position_basis}
\end{align}
where $R_{k \ell}$ have dimension eV. Note that the $R_{k \ell}$ here differ from those defined in Ref.~\cite{Catena:2019gfa} by a factor of $k/2 \pi$. Specifically, $R_{k \ell} = (k / 2 \pi) \sqrt{V} \bar{R}_{k \ell}$, where $\bar{R}_{k \ell}$ are defined in Ref.~\cite{Catena:2019gfa}.    
The $R_{k \ell}$ in Eq.~\eqref{eq:atomic_state_position_basis} satisfy,
\begin{align}
    \int dr \, r^2 R_{k' \ell}^* R_{k \ell} = 2 \pi \delta(k - k')
\end{align}
and are explicitly given by~\cite{Peng_2010},
\be
    R_{k \ell}(r; Z_F) & = \frac{2}{r} C_\ell \rho^{\ell + 1} e^{-i \rho} {}_{1}F_{1}(\ell + 1 - i \eta, 2 \ell + 2, 2 i \rho) 
    \ee
where we have defined 
\be
    C_l  =  \frac{2^\ell}{(2 \ell + 1)!}  e^{- \pi \eta / 2} \left| \Gamma(\ell + 1 + i \eta) \right| 
    ~~~,~~~\rho  = k r ~~~,~~~
    \eta  = -\frac{Z}{a_0 k} \, ,   
\ee
and ${}_{1}F_{1}(a, b, c)$ is the confluent hypergeometric function of the first kind. One common approximation for the $Z_F$ parameter~\cite{SuperCDMS:2018mne,Essig:2012yx,Essig:2017kqs} is to relate it to the binding energy of the state it was absorbed from, i.e.,
\begin{align}
    Z_F = n \times \sqrt{ - \frac{\omega_I}{13.6 \, \text{eV}} } \, ,
\end{align}
where $n$, $\omega_I$ are properties of the initial states used when calculating the transition matrix elements, $\mathcal{T}$.
With these conventions, the transition matrix elements become
\begin{align}
    \langle k \ell' m' | \mathcal{O} | n \ell m \rangle = \int d^3 \mathbf{x} R_{k' \ell '}^* Y_{\ell' m'}^* \left( \mathcal{O} \cdot \left( R_{n \ell} Y_{\ell m} \right) \right) \, .
\end{align}
In the next two subsections we derive explicit forms for the $\mathcal{O}$ needed to derive the atomic absorption rate for the magnetic dipole and electric dipole models studied in the main text.

\subsection{\texorpdfstring{$\mathcal{O} = v^i = - i \nabla^i / m_e$}{TEXT}}

The main identity needed is from Ref.~\cite{Catena:2019gfa},
\begin{align}
    \nabla^i \left( f(r) Y_{\ell m} \right) = \sum_{k = -1}^1 \sum_{q = -1}^1 \left( c^i(\ell, m, k, q) \frac{d f}{dr} + d^i(\ell, m, k, q) \frac{f}{r} \right) Y_{\ell + k, m + q} \, ,
    \label{eq:grad_equation}
\end{align}
where,
\begin{align}
    c^x(\ell, m, - 1, - 1) & = -i c^y(\ell, m, - 1, - 1) = - \frac{A_{--}}{2\sqrt{(2\ell - 1)(2\ell + 1)}} \\
    c^x(\ell, m, - 1, 1) & = i c^y(\ell, m,- 1, 1) = \frac{A_{-+}}{2\sqrt{(2\ell - 1)(2\ell + 1)}}\\
    c^x(\ell, m, 1, - 1) & = -i c^y(\ell, m, 1, - 1) = \frac{A_{+-}}{2\sqrt{(2\ell + 3)(2 \ell + 1)}} \\
    c^x(\ell, m, 1, 1) & = i c^y(\ell, m, 1, 1) = -\frac{A_{++}}{2\sqrt{(2\ell + 3)(2 \ell + 1)}} \\
    c^z(\ell, m, - 1, 0) & = \frac{A_{-0}}{\sqrt{2(2\ell - 1)(2\ell + 1)}} \\
    c^z(\ell, m, 1, 0) & = -\frac{A_{+0}}{\sqrt{2(2\ell + 3)(2\ell + 1)}}
\end{align}
\begin{align}
    d^x(\ell, m, - 1, - 1) & = -i d^y(\ell, m, - 1, - 1) = - \frac{(\ell + 1)A_{--}}{2\sqrt{(2\ell - 1)(2\ell + 1)}} \\
    d^x(\ell, m, - 1, 1) & = i d^y(\ell, m, - 1, 1) = \frac{(\ell + 1)A_{-+}}{2\sqrt{(2\ell - 1)(2\ell + 1)}}\\
    d^x(\ell, m, 1, - 1) & = -i d^y(\ell, m, 1, - 1) = - \frac{\ell A_{+-}}{2\sqrt{(2\ell + 3)(2 \ell + 1)}} \\
    d^x(\ell, m, 1, 1) & = i d^y(\ell, m, 1, 1) = \frac{\ell A_{++}}{2\sqrt{(2\ell + 3)(2 \ell + 1)}}\\
    d^z(\ell, m, - 1, 0) & = \frac{(\ell + 1)A_{-0}}{\sqrt{2(2\ell - 1)(2\ell + 1)}}\\
    d^z(\ell, m, 1, 0) & = \frac{\ell A_{+0}}{\sqrt{2(2\ell + 3)(2\ell + 1)}}
\end{align}
\begin{align}
    A_{++} & = \sqrt{(\ell + m + 1)(\ell + m + 2)} \\
    A_{+0} & = -\sqrt{2(\ell + m + 1)(\ell - m + 1)} \\
    A_{+-} & = \sqrt{(\ell - m + 1)(\ell - m + 2)} \\
    A_{-+} & = \sqrt{(\ell - m - 1)(\ell - m)} \\
    A_{-0} & = \sqrt{2(\ell + m)(\ell - m)} \\
    A_{--} & = \sqrt{(\ell + m - 1)(\ell + m)} \, .
\end{align}
All other $c^i$ and $d^i$ are zero, and we require that $\ell \geq 0, |m| \leq \ell, \ell + k \geq 0$ and $|m + q| \leq \ell + k$ for the spherical harmonic to have physical parameters. 

Using Eq.~\eqref{eq:grad_equation} the transition matrix element is given by,
\begin{align}
    \mathcal{T}_{v^i} & = \langle k \ell' m' | v^i | n \ell m \rangle \\
    & = -\frac{i}{m_e} \left[ c^i(\ell, m, \Delta \ell, \Delta m) \mathcal{I}_{nk\ell\ell'}^{0, 1} + d^i(\ell, m, \Delta \ell, \Delta m) \mathcal{I}_{nk\ell\ell'}^{1, 0} \right] \, ,
    \label{eq:atomic_T_vi}
\end{align}
where $\Delta \ell = \ell' - \ell, \Delta m = m' - m$, and we have defined the integral, $\mathcal{I}_{nk\ell\ell'}^{\alpha, \beta}$, as
\begin{align}
    \mathcal{I}_{nk\ell\ell'}^{\alpha, \beta} & \equiv \int dr \, r^{2} \,  \frac{R_{k \ell'}^*}{r^\alpha} \frac{d^\beta R_{n\ell}}{dr^\beta} \, .
    \label{eq:atomic_radial_int}
\end{align}
Note that the selection rules, $|\Delta \ell|  \leq 1$ and $|\Delta m| \leq 1 $ arise here from the finite nature of the sums over $k, q$.

\subsection{\texorpdfstring{$\mathcal{O} = v^i v^j = - \nabla^i \nabla^j / m_e^2$}{TEXT}}

To compute the second derivative of the initial state wave functions we simply need to use the identity in Eq.~\eqref{eq:grad_equation} twice,
\begin{align}
    \nabla^i \nabla^j \left( f(r) Y_{\ell m} \right) = \sum_{kk'qq'} \left( c^i c^j \frac{d^2 f}{dr^2} + \left( d^i c^j + d^j c^i \right) \frac{1}{r} \frac{d f}{dr} + d^j \left(d^i - c^i \right) \frac{f}{r^2}  \right) Y_{\ell + k + k', m + q + q'} \, ,
\end{align}
where it is understood that the $c, d$ superscripted with $i$ are evaluated at $(\ell, m, k, q)$, and the those superscripted with $j$ are evaluated at $(\ell + k, m + q, k', q')$.

Therefore the transition matrix element is given by,
\begin{align}
    \mathcal{T}_{v^iv^j} & = \langle k \ell' m' | v^i v^j | n \ell m \rangle \\
    & = - \frac{1}{m_e^2} \sum_{kq} \left( c^i c^j \mathcal{I}^{0, 2}_{nk\ell\ell'} + \left( d^i c^j + c^j d^i \right) \mathcal{I}^{1, 1}_{nk\ell\ell'} + d^j (d^i - c^i) \mathcal{I}^{2, 0}_{nk\ell\ell'} \right) \, ,
\end{align}
where $\mathcal{I}_{IF}^{\alpha, \beta}$ is defined in Eq.~\eqref{eq:atomic_radial_int}, and it is understood that the $c, d$ superscripted with $i$ are evaluated at $(\ell, m, k, q)$, and the those superscripted with $j$ are evaluated at $(\ell + k, m + q, \Delta \ell - k, \Delta m - q)$. The selection rules for this transition matrix element are $|\Delta \ell| \leq 2$ and $|\Delta m | \leq 2$.

\bibliographystyle{utphys3}
\bibliography{biblio}

\end{document}